\documentclass[12pt,preprint]{aastex}  

\newcommand{\mathEbreak}{E_{\mbox{\em\scriptsize break}}}
\newcommand{\mathEpeak}{E_{\mbox{\em\scriptsize peak}}}
\newcommand{\mathEFE}{E\!\cdot\!F_E}
\newcommand{\CB}{CBM}
\newcommand{\CLerrors}{90\% CL errors}
\newcommand{\fluenceCB}{$F_{\mbox{\scriptsize CB}}$}
\newcommand{\fluenceBand}{$F_{\mbox{\scriptsize Band}}$}
\newcommand{\fluenceCPL}{$F_{\mbox{\scriptsize CPL}}$}
\newcommand{\fluencemPL}{$F_{\mbox{\scriptsize mPL}}$}
\newcommand{\fluenceRHESSI}{$F_{\mbox{\scriptsize RHESSI}}$}
\newcommand{\fluenceHETE}{$F_{\mbox{\scriptsize HETE}}$}
\newcommand{\fluenceUlysses}{$F_{\mbox{\scriptsize Ulysses}}$}
\newcommand{\Ebreak}{$\mathEbreak$}
\newcommand{\Epeak}{$E_{\mbox{\em\scriptsize peak}}$}
\newcommand{\EFE}{$E\!\!\cdot\!\!F_E$}
\newcommand{\Ein}{E^{\mbox{\em\scriptsize in}}}
\newcommand{\Edet}{E^{\mbox{\em\scriptsize det}}}
\newcommand{\refeq}{eq.~\ref}
\newcommand{\fa}{\tablenotemark{a}}
\newcommand{\fb}{\tablenotemark{b}}
\newcommand{\PM}{$\pm$}
\newcommand{\keV}{\mbox{keV}}
\newcommand{\MeV}{\mbox{MeV}}
\newcommand{\phS}{\phm{-}}

\slugcomment{submitted to \apj, 2007-07-06}

\begin{document}


\title{
Observation of an unexpected hardening in the spectrum of GRB~021206
}


\author{C. Wigger\altaffilmark{1,2}, 
        O. Wigger\altaffilmark{2}, 
	E. Bellm\altaffilmark{3},
        W. Hajdas\altaffilmark{1}
}
\email{claudia.wigger@psi.ch}

\altaffiltext{1}{
    Paul Scherrer Institut,
    CH-5232 Villigen PSI,
    Switzerland
}
\altaffiltext{2}{
    Tellistrasse 9, 
    CH-5000 Aarau,
    Switzerland
}
\altaffiltext{3}{
    UC Berkeley Space Sciences Laboratory,
    7 Gauss Way, Berkeley, 
    CA 94720-7450, USA
}



\begin{abstract}
GRB 021206 is one of the brightest GRBs ever observed.
Its prompt emission, as measured by RHESSI,
shows an unexpected spectral feature.
The spectrum has a peak energy of about 700~\keV\
and can be described by a Band function up to 4.5~\MeV.
Above 4.5~\MeV, the spectrum hardens again, so that
the Band function fails to fit the whole RHESSI
energy range up to 17~\MeV.
Nor does the sum of a blackbody function plus a power law,
even though such
a function can describe a spectral hardening.
The cannonball model
on the other hand predicts such a hardening, 
and we found that it 
fits the spectrum of GRB 021206 perfectly. 
We also analysed other strong GRBs observed by RHESSI,
namely GRBs 020715, 021008, 030329, 030406, 030519B, 
031027, 031111. 
We found that all their spectra can
be fit by the cannonball model 
as well as by a Band function. 
\end{abstract}




\section{INTRODUCTION}
\label{sec:intro}

The exact mechanism which produces  $\gamma$-ray bursts (GRBs) 
has not yet been definitively established.
Their prompt $\gamma$-ray spectra can be used to distinguish 
between different models.
Several mathematical functions have been used for parametrizing
the prompt $\gamma$-ray emission.
Most commonly used is the empirical Band function \citep{Band93}, 
which is not motivated by a physical model.

There have been attempts to distinguish between spectral
models analysing the low energy part of the spectrum.
\citet{Ghirlanda2003},  \citet{Ryde2004}, and more recently
\citet{Ghirlanda2007} searched for blackbody components
in GRB spectra with varying degrees of success.
\citet{Preece2002}, using BATSE GRB spectra, 
tested the synchrotron shock model and conclude that
it ''does not account for the observed spectra 
during the GRB phase''. 


Spectral studies above the peak energy are rare,
one reason being the poor data quality because of lack of statistics.
Combining BATSE and EGRET spectra, 
\citet{Nature2003} report a high energy component for GRB 941017. 
They find a photon index of about 1.0 at energies above 5~\MeV.

In this paper we report a high energy 
component in GRB 021206 \citep{GCN_021206_IPN,GCN_021206_final},
observed with the Reuven Ramaty High Energy Solar Spectroscopic 
Imager RHESSI \citep{RHESSI}.
Having a peak energy of about 700~\keV, the spectrum of this burst
can be described by a Band function from 70~\keV\ up
to 4.5~\MeV, with a high energy photon index $\beta \approx 3.2$. 
Above 4.5~\MeV, the spectrum hardens again, and can be
described with a photon index  $\beta' \approx 2.2$.
This significant hardening around 4.5~\MeV\ can not
be described with a Band function. But it seems to 
differ from the spectral hardening in GRB 941017 as well.


There is one model that fits the entire RHESSI spectrum
of GRB 021206: the cannonball model \citep{Dar2004,Dado2002,Dado2003}.
The cannonball model
predicts a spectral hardening at several times the 
peak energy with a high energy photon index reaching $\beta \approx 2.1$. 

The question immediately arises whether the cannonball model
can improve our description of other GRB spectra.
The difference between the Band function and the cannonball model arises
only at the high energy part of the spectrum, where
data usually suffer from low statistics.
Therefore, we choose the strongest GRBs registered by RHESSI
in the years 2002 to 2004.
We find that they all  can be fit
by the cannonball model as well as by the Band function.

The outline of the paper is the following:
We first present shortly the instrument, the GRB selection,
the spectrum extraction, and the fit method (\S \ref{sec:observations}).
In the next section (\S \ref{sec:models}), 
many spectral functions are given.
In \S \ref{sec:results}, the fit results for 
GRB 020715, 
GRB 021008, 
GRB 021206, 
GRB 030329, 
GRB 030406, 
GRB 030519B, 
GRB 031027, 
and GRB 031111
are presented. The fits are discussed and, 
if possible, compared to other measurements.
The more general discussion, including
an outlook, follows in \S \ref{sec:discussion}.
We end with a short summary in \S \ref{sec:summary}.

\section{INSTRUMENT AND METHOD}
\label{sec:observations}

\subsection{Instrument}

RHESSI is a NASA Small Explorer mission
designed to study solar flares in hard X-rays and 
$\gamma$-rays \citep{RHESSI}.
It consists of two main parts:
an imaging system and the spectrometer
with nine germanium detectors \citep{spectrometer}.
The satellite always points towards the Sun
and rotates about its axis at $15\,$rpm. 
The Ge detectors are arranged in a plane perpendicular 
to this axis. 

The shape of the detectors is cylindrical with a
height of $\approx 8.5\,$cm and diameter of $\approx 7.1\,$cm,
and they
are segmented into a thin front ($\approx 1.5\,$cm) 
and a thick rear segment ($\approx 7\,$cm).
Since the shielding of the rear segments is minimal,
photons with more than about 25~\keV\ can enter
from the side.
Above about 50--80~\keV, photons from any
direction can be observed.
Each detected photon is time- and energy-tagged from
3~\keV\ to 2.8~\MeV\ (front segments) 
or from 20~\keV\ to 17~\MeV\ (rear segments).
The energy resolution is $\approx 3$~\keV\ at 1~\MeV, and the
time resolution is 1~$\mu$s.

The effective area for GRB detection
depends on the incident photon energy $E$
and the angle between the GRB direction
and the RHESSI axis, the incoming angle $\theta$.
Over a wide range of $E$ and $\theta$, the
effective area is around 150$\,$cm$^2$.
The sensitivity drops rapidly at energies below
$\approx 50$~\keV.

\subsection{GRB selection}
\begin{figure}
\epsscale{0.52} 
\plotone{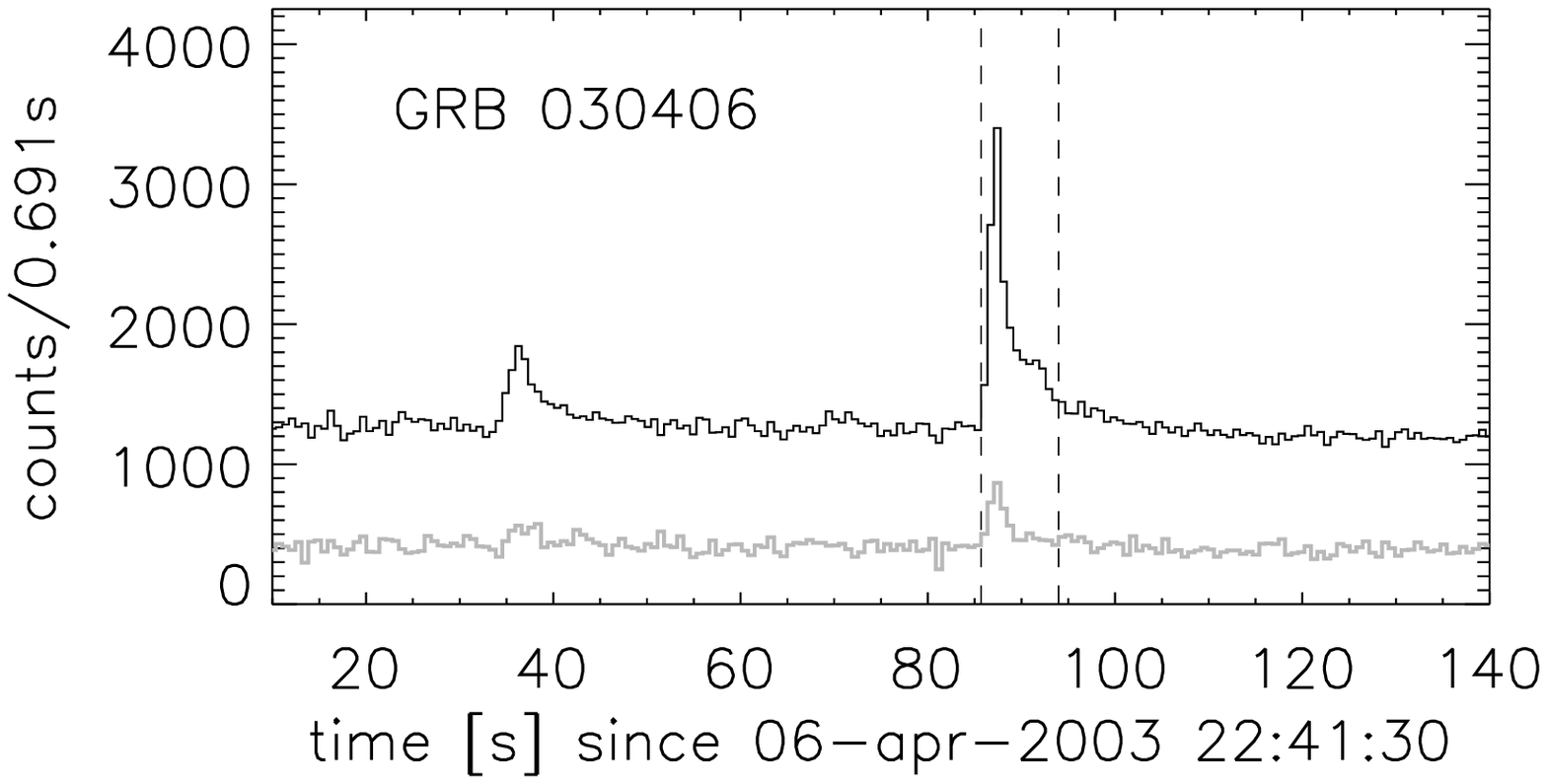} 
\plotone{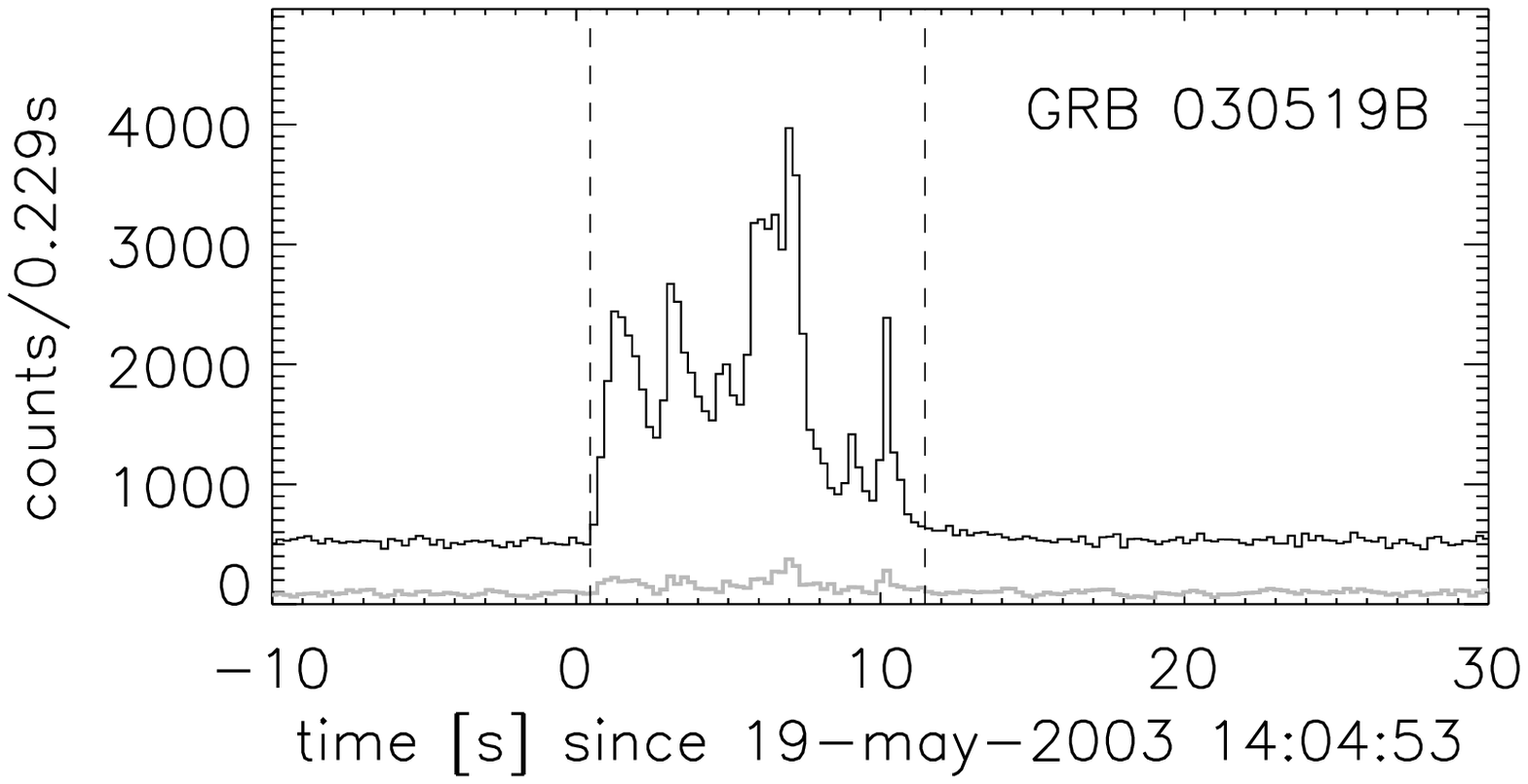} 
\plotone{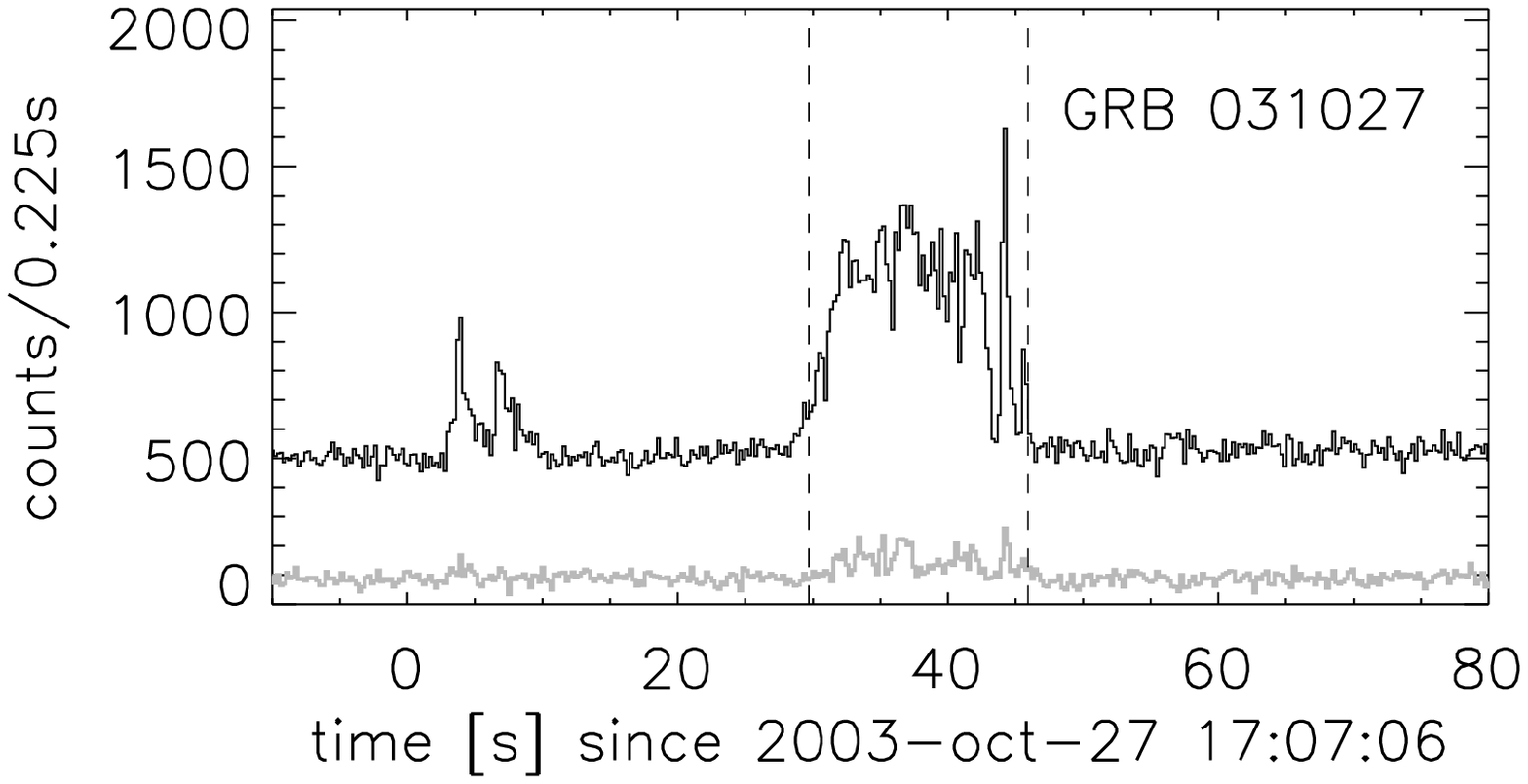} 
\plotone{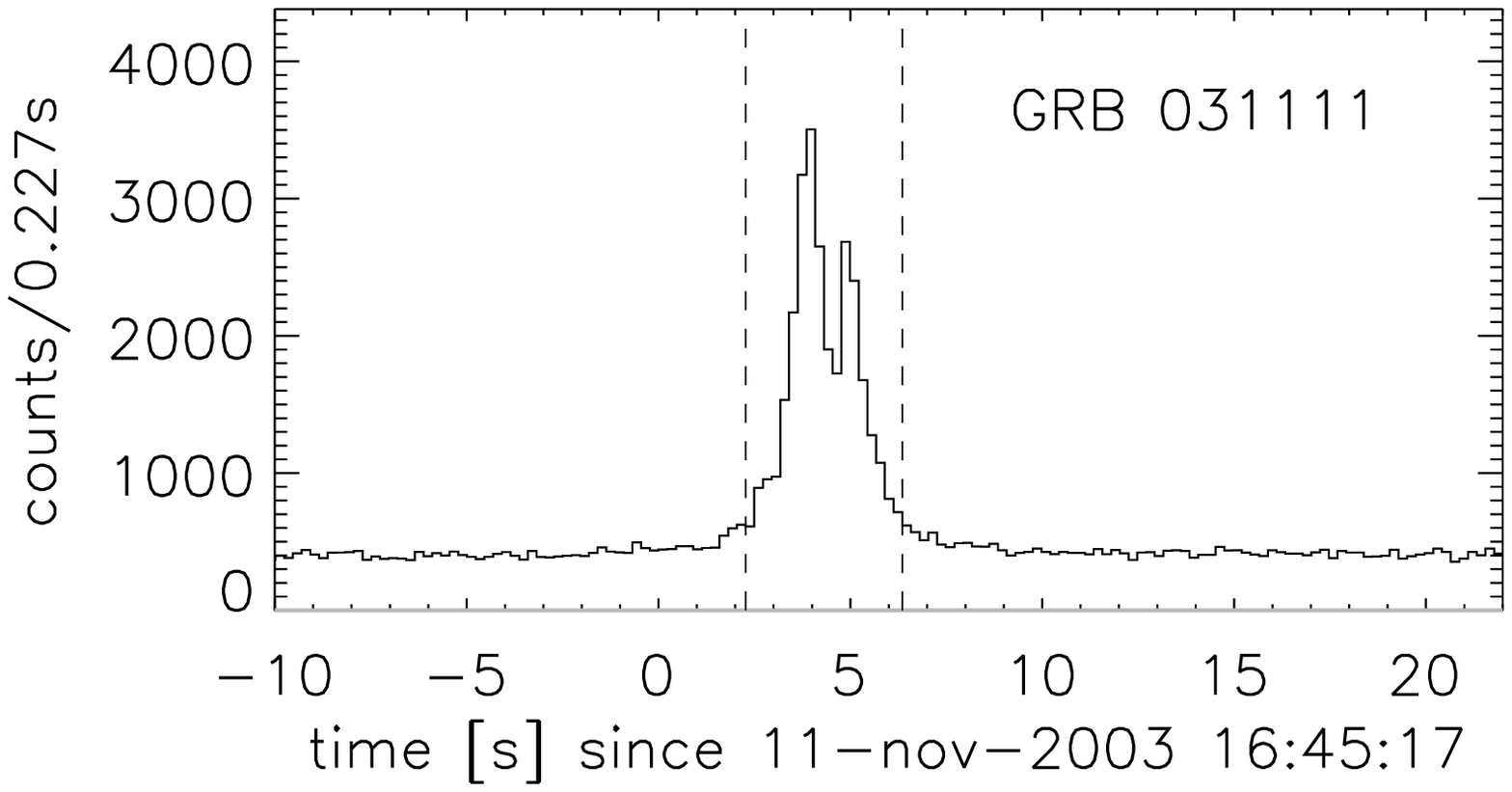} 
\caption{lightcurves
  for the energy band 20~\keV\ to 3~\MeV;
  black: rear detectors,
  grey: front detectors;
  broken vertical lines:
  time intervals used for spectral analysis.
  \label{fig:ltc_GRBsII}}
\end{figure}

For this study, we need well observed GRB spectra.
We chose GRBs from the years 2002 to 2004, because
radiation damage starts to play a role in 2005.
The selected GRBs have to be localized by other observations of
the same GRB (RHESSI can not measure the incoming angle),
because $\theta$ enters into the simulation of the response function.
A further requirement was the availability of good background data. 
And finally, the data storing mode
(`rear decimation' for onboard memory saving)
is not allowed to change during the entire GRB and background time interval.
Of all the GRBs meeting these criteria, we chose eight with
the best signal-to-background ratio,
listed in Table~\ref{tab:GRBs}
along with their incoming angle $\theta$ and
the time intervals used.
The lightcurves of these bursts are shown in
Figs.\ \ref{fig:ltc_GRBsI} and \ref{fig:ltc_GRBsII}.

\subsection{Preparation and fit of RHESSI spectra}
\label{sec:fit}
The method of analysing RHESSI GRB spectra
will be described in detail in a separate article
(E.\ Bellm, C.\ Wigger et al., in preparation).

For each GRB and detector segment, 
the total spectrum (GRB plus background) 
during the burst was extracted, 
as well as the background spectra during two time intervals
before and after the burst.
The background was linearly (sometimes quadratically) interpolated
and subtracted.
The exact time intervals are listed in Table \ref{tab:GRBs}.
Then we added all rear and all front segments,
except for detector \#2 which
is slightly damaged and has a bad energy resolution.
Since all GRBs in this study are strong,
the observational errors 
are dominated by the statistical error of the GRB
counts, not of the background.

We simulate RHESSI using GEANT3 \citep{Geant3}.
Knowing the direction of the GRB from other instruments, we simulate
RHESSI's response to photons coming from angle $\theta$.
The energy of the incoming photons is simulated as
a power law spectrum (i.e.\ $dN/dE \propto E^{-\gamma_{sim}}$)
with typically $\gamma_{sim} = 2\,$. 
%
%
This power law simulation is only a rough approximation
and is not not intended to represent the intrinsic GRB spectrum, 
but instead provides simulated data representing RHESSI's conversion of
photons to counts.
The true GRB source spectrum is determined via weight factors 
for the resulting simulated count spectrum, as described below.
The upper energy limit of the simulated photon spectrum 
is typically 30 MeV, in the case of GRB 021206 even 40 MeV or 50 MeV.
This is important,  because an incoming
photon of e.g.\ 25 MeV may well make a signal of 15 MeV.
Rotation angles are generated uniformly,
i.e.\ we compute a RHESSI-spin averaged response function.
Since the detector arrangement
shows an approximate 120 degree symmetry,
the averaging gives good results as long
as the analysed time interval is at least one third of
the rotation period ($T_{rot} = 4\,$s). 
This was also confirmed by tests.

The output of the simulation is
an event list, or rather a hit list,
consisting of all signals registered
in the Ge detectors.
The simulated hit list, having $N_{s}$
entries indexed by the letter $l$, contains the deposited energy
($\Edet_l$) as well as the initial photon energy ($\Ein_l$).
The measured hit list contains only the observed energy. 

For spectral fitting, the observed energy histogram
is compared with a histogram accumulated from the simulated
hit list.
More precisely: 
The measured histogram can be represented by
a $k$-element vector $\vec M$ with errors $\vec\sigma_M$, and 
energy boundaries $E^b_0$, $E^b_1$, $E^b_2$,
..., $E^b_k$. 
We normalize the histogram $\vec M$ to the total
number of counts in the fit range: 
$\vec{m} = \vec{M}/C_M $
and $\vec{\sigma}_m = \vec{\sigma}_M /C_M $, where 
$C_M = \sum_{i \in I} M_i $ and the sum goes only over the
bins included in the fit, i.e.\ 
$I = \{i \mid \mbox{bin } i \mbox{ is included in the fit} \}$.
The 'theoretical' histogram $\vec{S}$ is accumulated from the 
simulated hit list. 
Each entry is weighted with a factor in order to scale 
from the simulated power law 
to the probability density which would be expected, 
had we actually simulated the GRB source spectrum dN/dE.
The $j$th 
bin contains therefore the weighted sum of all simulated hits
with $\Edet_l$ belonging to that bin, i.e.: 
\begin{equation}
   S_j = \sum_{l \in L} w_l
\end{equation}
where $L = \{l \mid E^b_{j-1} \leq \Edet_l < E^b_j\}$
and 
\begin{equation}
   w_l = \left( \frac{\Ein_l}{E_{piv}} \right)^{\gamma_{sim}} \cdot
       \frac{dN}{dE}(\Ein_l)   \label{eq:weight}
\end{equation}
%
%
The first factor in eq.\ \ref{eq:weight} accounts
for the spectrum assumed in the simulation and 
the energy $E_{piv}$ is an arbitrary normalisation.
The second factor accounts for
the spectrum of the incoming GRB photons. 
Possible parametrisations
of $dN/dE$ are given below in \S \ref{sec:models}.  
If the GRB spectrum had 
the same shape as the simulated one,
 i.e.\ if
$ dN/dE = (E/E_{piv})^{-\gamma_{sim}}$, 
the weights would all be 1 .
This  method of using weight factors when filling a histogram
is common in particle physics, see e.g.\ \citet{Barlow93}. 
The statistical error of the 'theoretical'
histogram $\vec{S}$ is $\sigma^2_{S_j} = \sum_{l \in L} w_l^2 $
\citep[\S 6 of][]{Barlow93}.
%
As in the case of the measured histogram,
the histogram $\vec{S}$
is normalised:
$\vec{s} = \vec{S}/C_S $
and $\vec{\sigma}_s =\vec{\sigma}_S /C_S  $ with 
$C_S = \sum_{i \in I} S_i $. 

The parameters of the histogram $\vec{S}$ are varied
until the minimum of
\begin{equation}
   \chi^2 = \sum_{i \in I} \frac{(m_i - f s_i)^2}
      {\sigma_{m_i}^2 + \sigma_{s_i}^2}
\end{equation}
is found. The factor $f$ accounts for the
normalisation between measured and simulated histogram.
It is expected to be almost 1, but
should be treated as a free fit parameter.
For each fit iteration, the histogram $\vec{S}$ is recalculated
with different weights (eq.\ \ref{eq:weight}).  

Since the simulated hit list contains many more
photons than the measured spectrum, 
we used the approximation $\sigma_{m_i}^2 + \sigma_{s_i}^2 \approx 
\sigma_{m_i}^2$ while fitting.
But it was always checked that
the statistical error from the measurement is
dominant.

It is possible to create a response matrix from our 
simulations and perform spectral fits via forward-fitting,
as in  XSPEC.  
In any case, our weighted histogram method gives equivalent
fits to response matrices which are simulated directly 
(by EB; see E.\ Bellm, C.\ Wigger et al., in preparation).



\subsection{Systematic effects}
\label{sec:systematics}
At low energies, a small deviation of our RHESSI mass model 
from the true amount of material 
can make a considerable difference in the number of 
observed photons.
For $\theta\approx 90$ degrees,
this should be a small problem because the 
lateral shielding is thin. But for $\theta$
from 10 to 50 degrees this is an 
issue, and less prominently also
from 130 to 160 degrees.


Simulation quality also gets better with higher energy.
This is fortunate for the current analysis which relies on
high-energy properties of GRB spectra.

\section{SPECTRAL MODELS}
\label{sec:models}

Let $dN/dE$ be the number of GRB photons per energy bin.
The peak energy \Epeak\ is defined as the energy for which
$\mathEFE = dN/dE \, E^2$ is maximal. 
The spectrum in the \EFE\ representation has at least one maximum because
the total emitted energy must be finite: $\int_0^\infty dN/dE\,E\, dE < \infty$.
Many instruments can see such a maximum \Epeak\ within their energy range.

Different mathematical functions, sometimes called models,
can describe such a shape. A collection is presented 
in this section. 



The simplest spectral function is a power law ({\bf PL}):
  \begin{equation} 
    \frac{dN}{dE} = A \left( \frac{E_{piv}}{E} \right)^\gamma
    \label{eq:PL}
  \end{equation}
where $E_{piv}$ is a normalization energy, e.g. $E_{piv}=100$~\keV.
The PL has no peak energy.
It rarely fits a GRB spectrum over the entire observed
energy range, but it is often useful for a limited energy band.
Indeed, every spectrum can be fit by several joined PLs.

One simple way to account for a spectral
softening and a peak energy is the cut off power law ({\bf CPL}):
\begin{equation}
  \frac{dN}{dE} = A \left( \frac{E_0}{E} \right)^\alpha e^{-E/E_0}
  \label{eq:CPL}
\end{equation}
If $\alpha < 2.0$ then $\mathEpeak = E_0\, (2-\alpha)$ .

Another way to account for a spectral break is the
broken power law ({\bf BPL}) consisting of two joined PLs
\begin{equation}
  \frac{dN}{dE} = 
  \left\{
    \begin{array}{ll}
      A \left( \frac{E_b}{E} \right)^\alpha & \mbox{ if } E\leq E_{b} \\
      A \left( \frac{E_b}{E} \right)^\beta  & \mbox{ if } E\geq E_{b}.
    \end{array}
  \right.
  \label{eq:BPL}
\end{equation}
If $\alpha < 2.0$ and $\beta > 2.0$, then $\mathEpeak = E_b$ .
This function is not continuously differentiable.

A smooth transition between the two power laws 
is realized by the empirical {\bf Band} function \citep{Band93}.
This is a smooth composition of a CPL for low energies and
a PL for high energies:
\begin{equation}
  \frac{dN}{dE} =
  \left\{
    \begin{array}{ll}
       A \left( \frac{E_{piv}}{E} \right)^\alpha e^{-E/E_0}  & \mbox{ if } E \leq \mathEbreak \\ 
       B \left( \frac{E_{piv}}{E} \right)^\beta 	     & \mbox{ if } E \geq \mathEbreak
    \end{array}
  \right.
  \label{eq:Band}
\end{equation}
where
\[
   \mathEbreak =  E_0 (\beta-\alpha)
\]
and
\[
   B = A \left(\frac{E_0}{E_{piv}} (\beta-\alpha)\right)^{\beta-\alpha}  
     e^{-(\beta-\alpha)}
\]
Again, $E_{piv}$ is a normalization energy, e.g. $E_{piv}=100$~\keV.
If $\alpha < 2.0$ and $\beta > 2.0$, then $\mathEpeak = E_0\, (2-\alpha)$.
If $\beta \longrightarrow \infty$ or if \Ebreak\ lies 
at the upper limit of the observed energy range, 
then the Band function turns into a CPL.
As already pointed out by \citet{Preece2000}, section 3.3.1.,
the low energy photon index, the curvature of the spectrum,
and its peak energy are represented with only two parameters,
$\alpha$ and $E_0$.\footnote{The smoothly broken power law model (SBPL, 
see \citet{Preece2000,BATSEcatalog})
would account for this problem with an additional parameter, 
but we do not use it here, because it can not fit
a spectral hardening at high energies.}

Sometimes a blackbody spectrum plus power law
is used for spectral fitting, 
see e.g.\ \citet{Ryde2004,Ghirlanda2007,SMcBreen2006}. 
We will call this the {\bf BBPL}:
\begin{equation}
  \frac{dN}{dE} =  A  \frac{(E/(kT))^2}{\exp(E/(kT))-1} + 
       A\,b\, \left( \frac{E_{piv}}{E} \right)^\alpha  
  \label{eq:BBPL}
\end{equation}  
We choose $E_{piv} = 3.92\, kT$, the
peak energy of the blackbody component.
The BBPL function can fit a spectral hardening at high 
energies.

When fitting with the BBPL model, it is often found that
the PL component does not fit simultaneously at low
and at high energies. This is also mentioned by \citet{Ryde2004}.
We therefore invented a blackbody plus modified power law
({\bf BBmPL}):
\begin{eqnarray}
  \label{eq:BBmPL}
  \frac{dN}{dE} =  A  \frac{(E/(kT))^2}{\exp(E/(kT))-1}\, + \\
                   A \, b \, (1-e^{-E/E_0}) \left( \frac{E_0}{E} \right)^\alpha \nonumber
\end{eqnarray}
We choose again $E_0=3.92\, kT$.
The modification of the power law component was borrowed from
the cannonball model (see next). The BBmPL function can 
describe a spectral hardening at high energies.


The cannonball model {\bf \CB} 
\citep{Dar2004,Dado2002,Dado2003}
makes a prediction for the
spectral shape of the prompt GRB emission.
It consists of a CPL and a modified power law:
\begin{eqnarray}
  \label{eq:CB}
  \frac{dN}{dE} = A \left( \frac{T}{E} \right)^\alpha  e^{-E/T} +  \\
	          A\, b\, \left( \frac{T}{E} \right)^\beta \left( 1- e^{-E/T}\right)  \nonumber 
\end{eqnarray}
according to \citet{Dar2004},  eq.~47, or \citet{Dado2004}, eq.~13.
The theoretically expected values are
$\alpha \approx 1.0$ and 
$\beta \approx 2.1$. 
The \CB\ function \refeq{eq:CB} applies strictly speaking 
only to the spectrum caused by a single cannonball,
i.e.\ for every single peak of a GRB.

It is often observed that the peak energy \Epeak\ is
a more stable fit parameter than the parameter $E_0$ in the Band function
(\refeq{eq:Band}) or in the CPL (\refeq{eq:CPL}).
Therefore, we use 
$E_p = E_0\, (2-\alpha)$ as a fit parameter. Similarly, we use
$T_p = T\, (2-\alpha)$ as a fit parameter in the case of 
\CB\ (\refeq{eq:CB}).

A word about fitting \CB:
For the high energy part, it has two parameters, whereas
the Band function has only one.
Already when fitting the Band function, it is
often observed that the high energy power law index
is poorly constrained, because the high energy data
tend to have large statistical errors.
This is even worse for \CB\ with two
high energy parameters. 
%
It often helps to freeze the parameter $\beta$
at its theoretical value of $\beta=2.1$
in order to make the fit converge.

\section{FIT RESULTS AND FIT DISCUSSIONS}
\label{sec:results}

%

The spectral models used and the $\chi^2$ of
the fits are listed in Table~\ref{tab:chi2} for all eight GRBs. 
For \CB\ and Band function, the fitted
parameters are listed in Tables~\ref{tab:CB_pars}
and \ref{tab:Band_pars}, respectively.
Throughout this article, all errors are symmetric $1\sigma$ errors
if not stated otherwise.

\begin{figure}
\epsscale{1}
\plotone{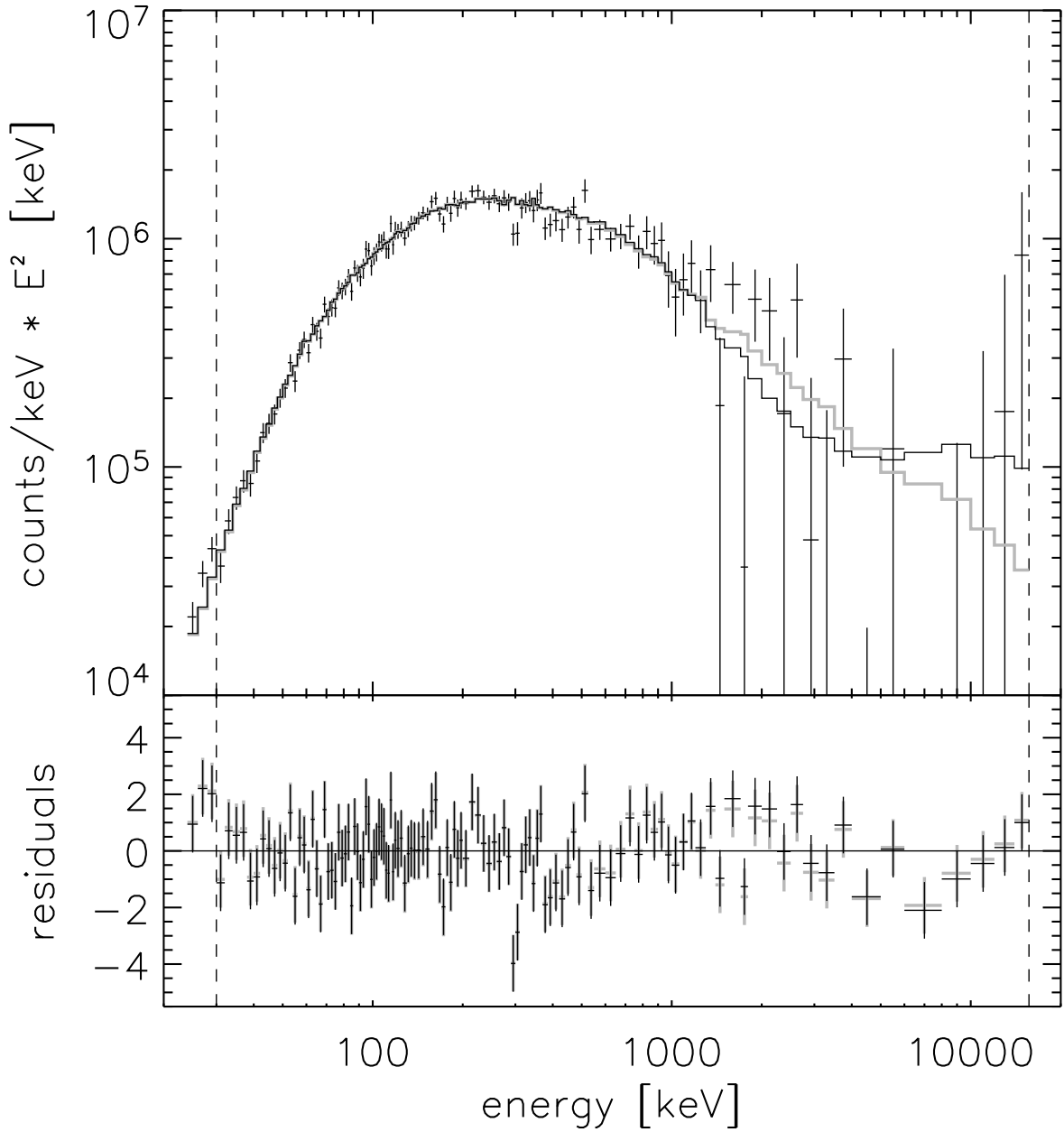} 
\caption{rear spectrum of GRB 020715;
   error bars: photon counts after background subtraction;
   black histogram: \CB\ fit (\refeq{eq:CB});
   grey histogram:  Band function fit (\refeq{eq:Band});
   vertical broken lines: energy range used for fitting;
   bottom: residuals of \CB\ (black) and Band (grey) fit.
  \label{fig:020715}}  
\end{figure}

\begin{figure}
\plotone{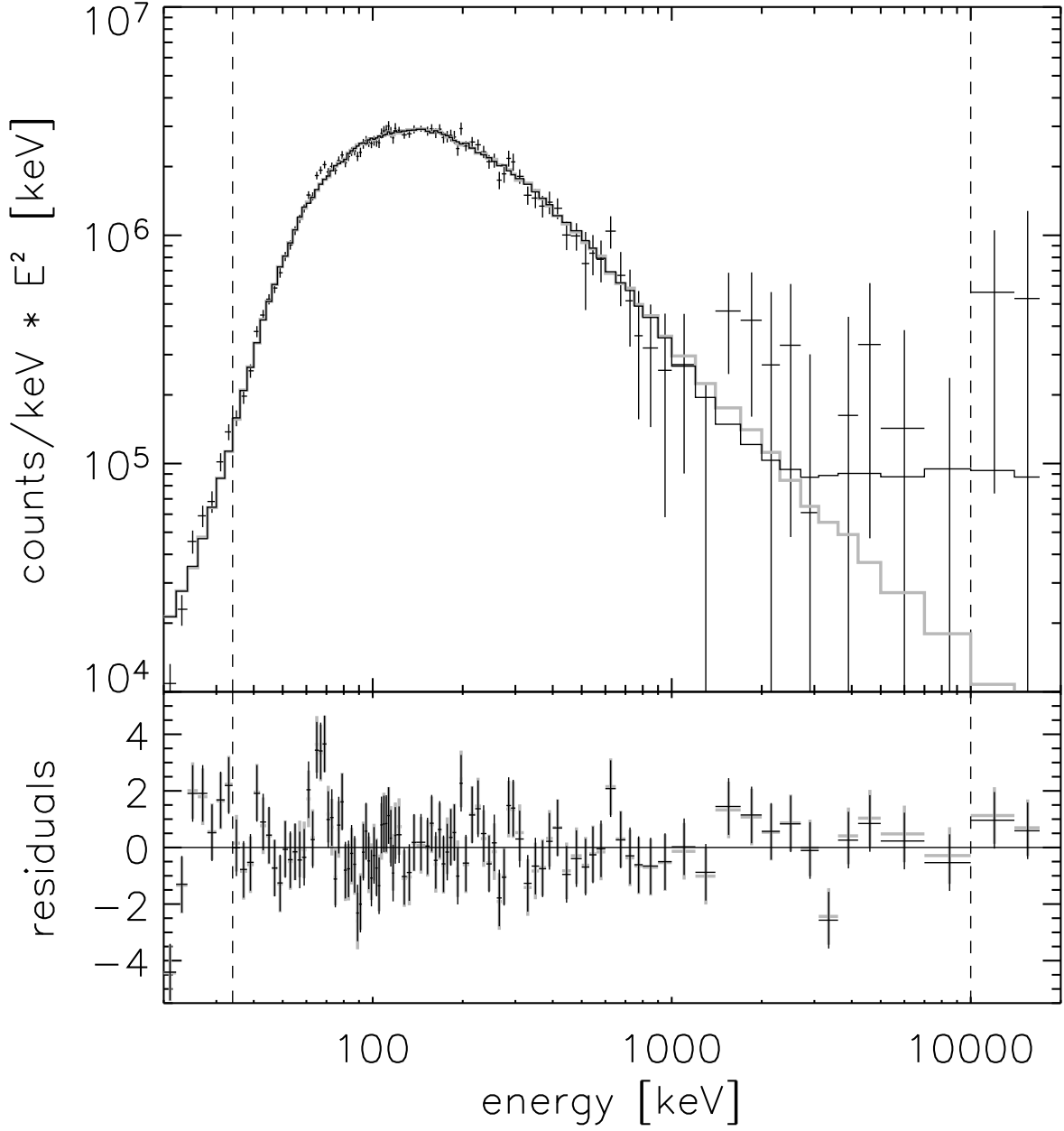} 
\caption{rear spectrum of GRB 030329, first peak;
   explanations see caption of Fig.~\ref{fig:020715}.
   \label{fig:030329_p1}}  
\end{figure}

\begin{figure}
\plotone{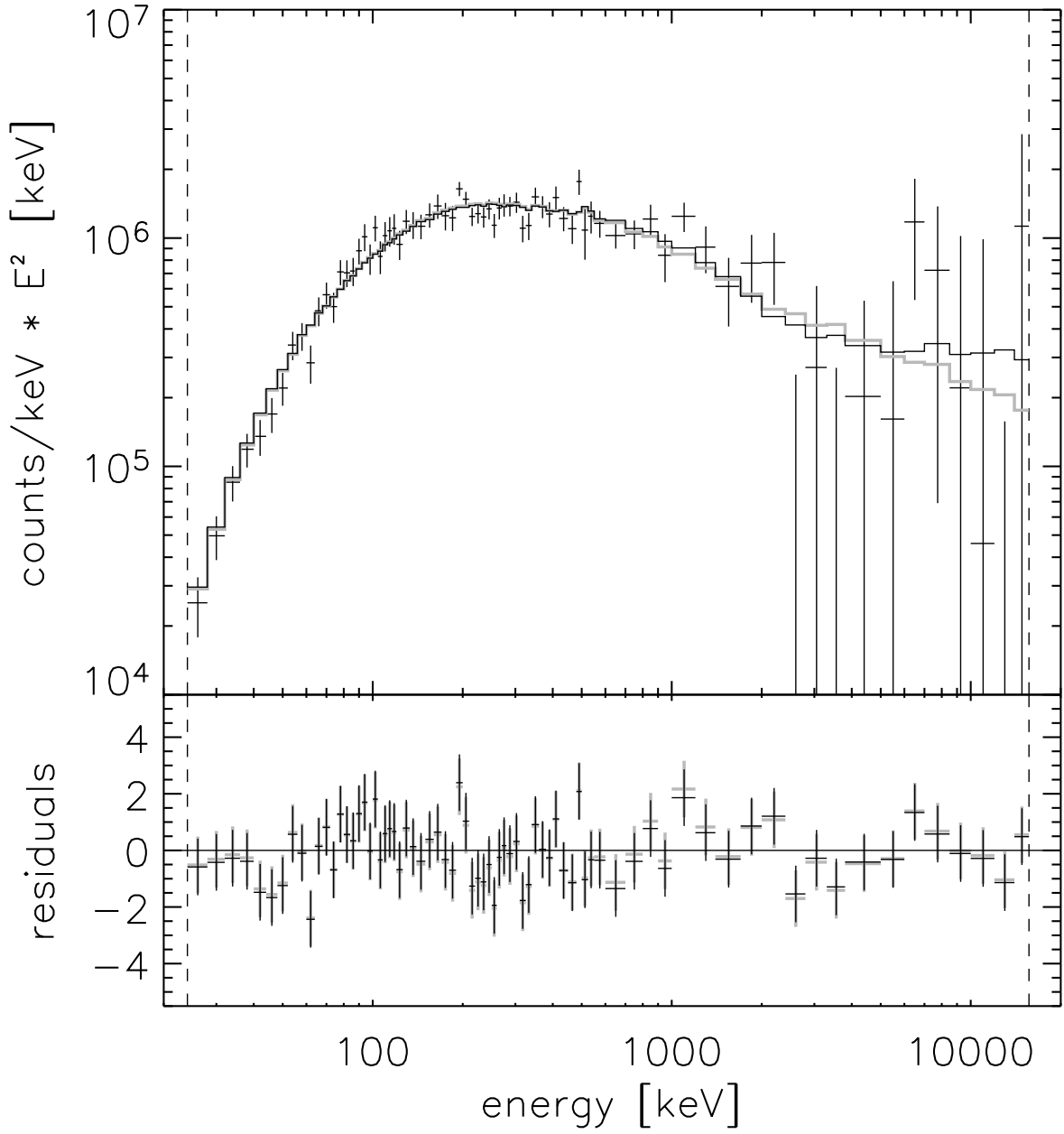} 
\caption{rear spectrum of GRB 030406;
   explanations see caption of Fig.~\ref{fig:020715}.
  \label{fig:030406}}  
\end{figure}

\begin{figure}
\plotone{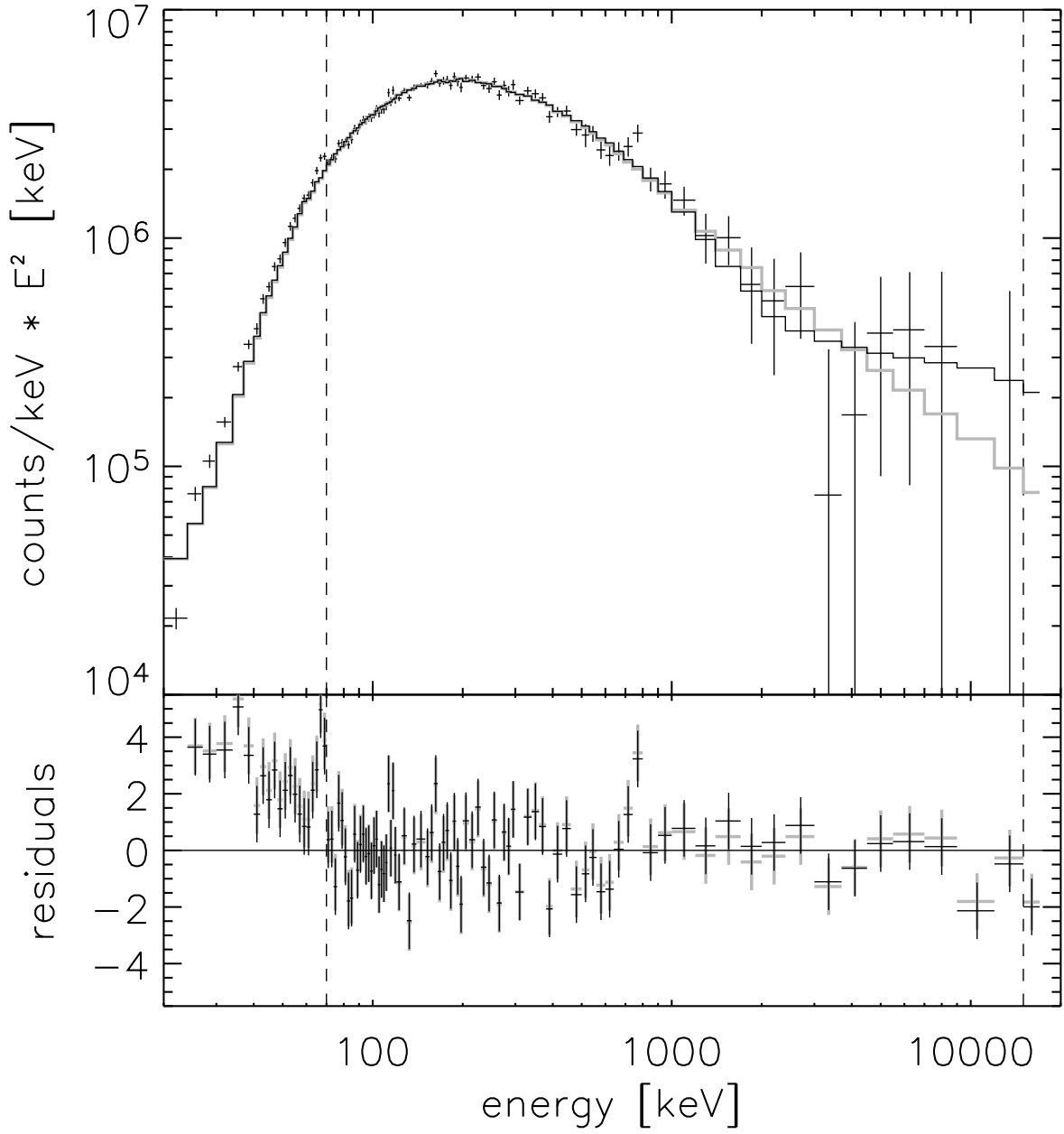} 
\caption{rear spectrum of GRB 030519B;
   explanations see caption of Fig.~\ref{fig:020715}.
  \label{fig:030519B}}  
\end{figure}

\begin{figure}
\plotone{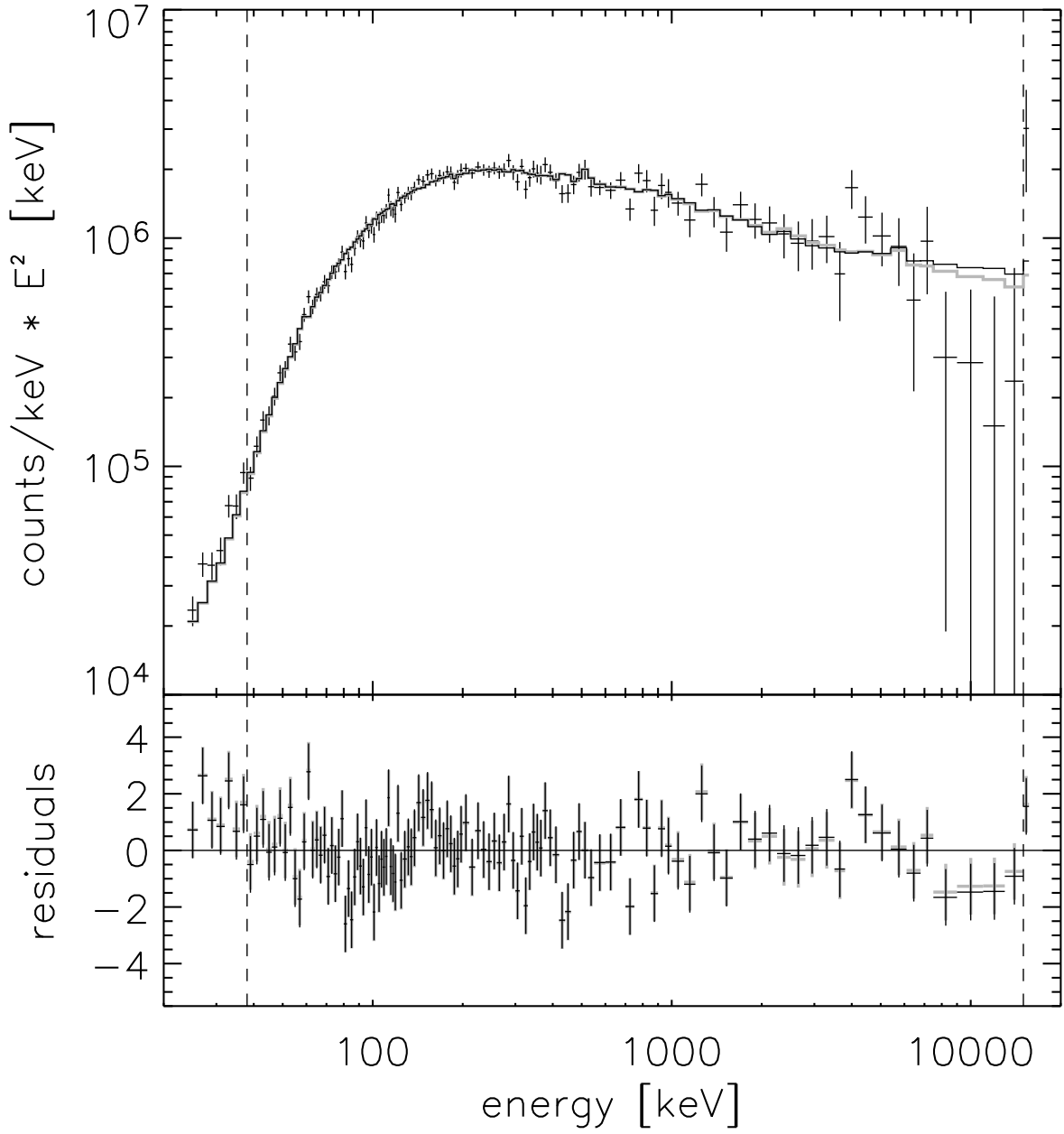} 
\caption{rear spectrum of GRB 031111;
   explanations see caption of Fig.~\ref{fig:020715}.
  \label{fig:031111}}  
\end{figure}

The measured spectra together with the \CB\ and Band fits
are shown in Figs.~\ref{fig:020715} to \ref{fig:031111}.
Note that we plot energy$^2 \,\cdot\,$counts/keV versus energy.
The difference to a deconvolved \EFE\ distribution is
discernible e.g.\ in the drop of counts
towards lower energies in our representation.
The statistical scatter from the limited number
of simulated events is sometimes visible as a little roughness of
the simulated spectra.\footnote{
In the case of e.g.\ GRB 021206, rear (see Fig.~\ref{fig:spec_rear}), 
the mean measured error between 4 and 12 MeV is  
$0.65 \cdot 10^6\,$counts$\cdot$keV,
whereas the mean scatter of the simulated histogram
is $0.25 \cdot 10^6\,$counts$\cdot$keV. For the other
GRBs with less observed photons and therefore larger measurement errors,  
the statistical error of the
simulation is even more negligible.}

From the fit parameters obtained for the
\CB\ and the Band function, we calculate the fluences
for various energy intervals.
They are listed in Table~\ref{tab:fl}.
%
%
The error of the fluence is dominated by systematics,
e.g.\  because we do not know the exact active volume 
of the single detector segments. 
We estimate the systematic error to be of order 5\%,
whereas the statistical error 
is of order 1\%.
Note also, that the two fluences
obtained by fitting \CB\ and Band function
are nearly equal.

\subsection{GRB 020715}
The lightcurve is shown in Fig.~\ref{fig:ltc_GRBsI} (top) and 
the spectrum is shown in Fig.~\ref{fig:020715}.
Two bins from 290~\keV\ to 310~\keV\ (a background line) 
and one bin from 500~\keV\ to 525~\keV\ (the 511~\keV\ line) 
were omitted in the fit 
because they would dominate $\chi^2$.
The best fit 
is a Band function, but the \CB\ also fits well.


\subsection{GRB 021008}
\begin{figure}
\plotone{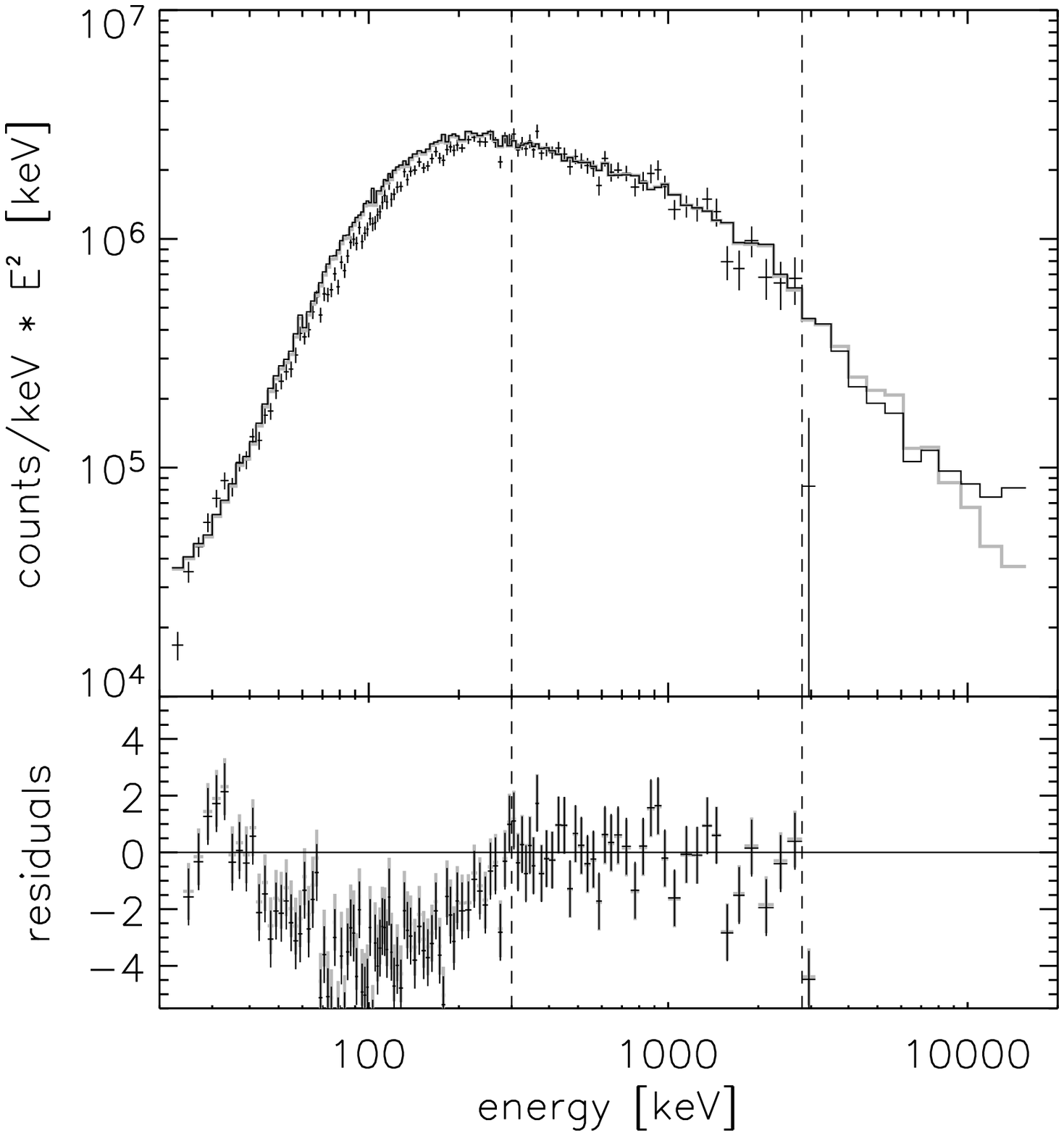} 
\caption{front spectrum of GRB 021008;
   explanations see caption of Fig.~\ref{fig:020715};
  the same set of parameters 
  (eqs.~\ref{eq:CB021008} and \ref{eq:Band021008}) 
  is used for this plot and Fig.~\ref{fig:R021008}.
  \label{fig:F021008}}  
\end{figure}

\begin{figure}
\plotone{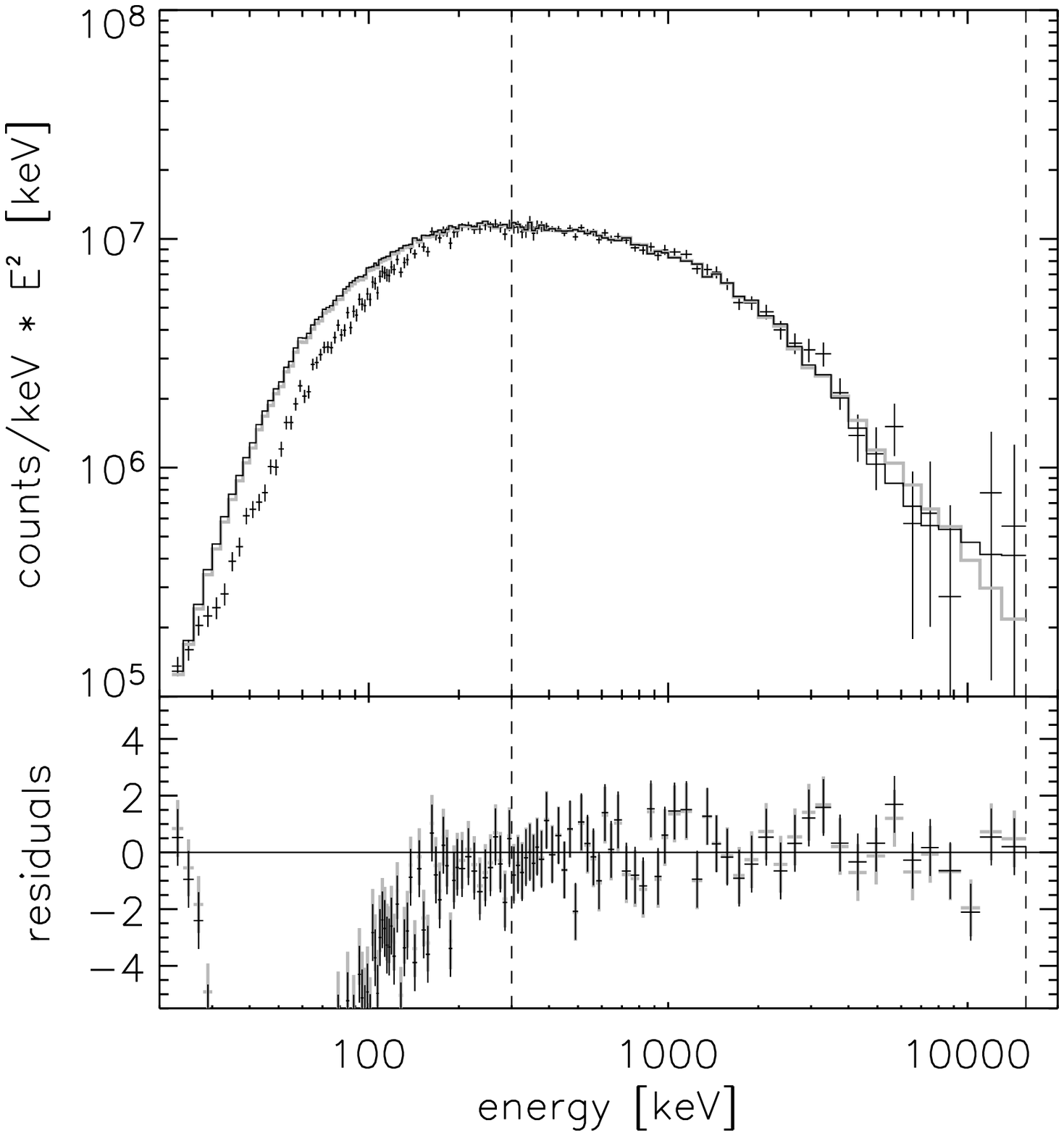} 
\caption{rear spectrum of GRB 021008;
   explanations see caption of Fig.~\ref{fig:020715}.
  the same set of parameters
  (eqs.~\ref{eq:CB021008} and \ref{eq:Band021008})
  is used for this plot and Fig.~\ref{fig:F021008}.
  \label{fig:R021008}}  
\end{figure}

Coming from a direction about 50 degrees from the Sun,
this GRB deposited photons not only in rear
detectors but also in the front detectors, 
as can be seen from the lightcurve in Fig.~\ref{fig:ltc_GRBsI}.
The front spectrum is shown in Fig.~\ref{fig:F021008}
and the rear in Fig.~\ref{fig:R021008}.

\paragraph{Fit}
We had difficulties to fit front and rear spectra consistently
below 300~\keV. We therefore chose 300~\keV\ as the lower energy bound. 
Band function and \CB\ give the best fits.

We fitted front and rear segments separately, as well as jointly.
The results of the joint fit are shown in 
Figs.~\ref{fig:F021008} and \ref{fig:R021008} .
For the joint \CB\ fit we find the 90\% confidence level (CL) errors:
\begin{eqnarray}
   T_p    &   =    & 641   \; ^{+54}_{-53} \mbox{ keV} \nonumber	        \\
   \alpha &   =    & 1.523 \; ^{+0.099}_{-0.098}	    \label{eq:CB021008} \\
   \beta  & \equiv & 2.1  			    \nonumber	        \\
    b     &   =    & 0.0198\; ^{+0.0129}_{-0.0137}  \nonumber
\end{eqnarray}
The total $\chi^2$ is $71.2$  for $n_{DoF}= 74$.
The Band function fits marginally better,
$\chi^2= 70.0$ for $n_{DoF}= 74$ 
and its parameters are (\CLerrors):
\begin{eqnarray}
   E_p    &=& 677  \; ^{+51}_{-66}       \mbox{ \keV} \nonumber \\
   \alpha &=& 1.493\; ^{+0.104}_{-0.104} \label{eq:Band021008} \\
   \beta  &=& 3.73 \; ^{+0.48}_{-0.38}	 \nonumber
\end{eqnarray}

\paragraph{Discussion}
We do not well understand the  
spectrum below 300~\keV.
Both fits, the \CB\ and the Band function,
overestimate the counts below 300~\keV. This can be a hint that
the GRB spectrum hardens below 300~\keV.
We find functions that fit the front and
the rear data from 40~\keV\ to 400~\keV\ individually,
but they do not agree. 

One possible explanation is the GRB incoming angle of 
about 50 degrees at which the
GRB photons pass through a certain amount
of material before reaching the detectors.
Our GEANT simulation tries to take that into account, but it
is probably not perfect, and maybe the averaging over all rotation
angles is a bad assumption for this short GRB pulse.


Another difficulty for this GRB is its background.
For the single rear segments, the background at low
energies (below $\approx 120$~\keV)
strongly depends on the rotation angle of RHESSI.
We did our best to take this into account, 
but maybe did not succeed completely.

\subsection{GRB 021206}

\label{sec:result021206}
GRB 021206 is famous for its claimed
polarization \citep{CB03}, which however
turned out to be an artefact 
\citep[see][]{RF04,PSI2004,cw_rome}.
This GRB was also studied by \citet{Boggs2004} 
to probe quantum gravity.

\begin{figure}
\epsscale{0.52} 
\plotone{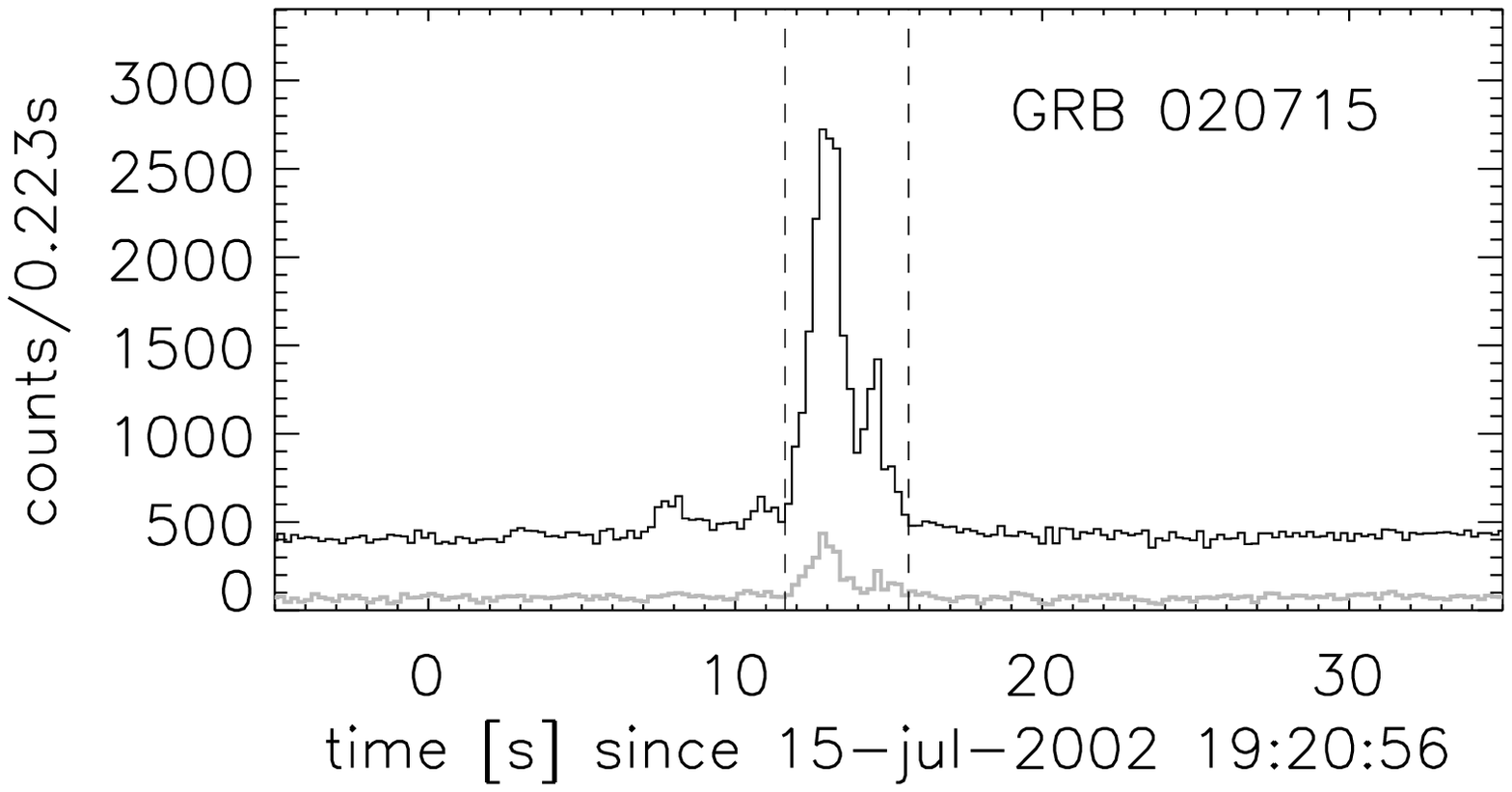} 
\plotone{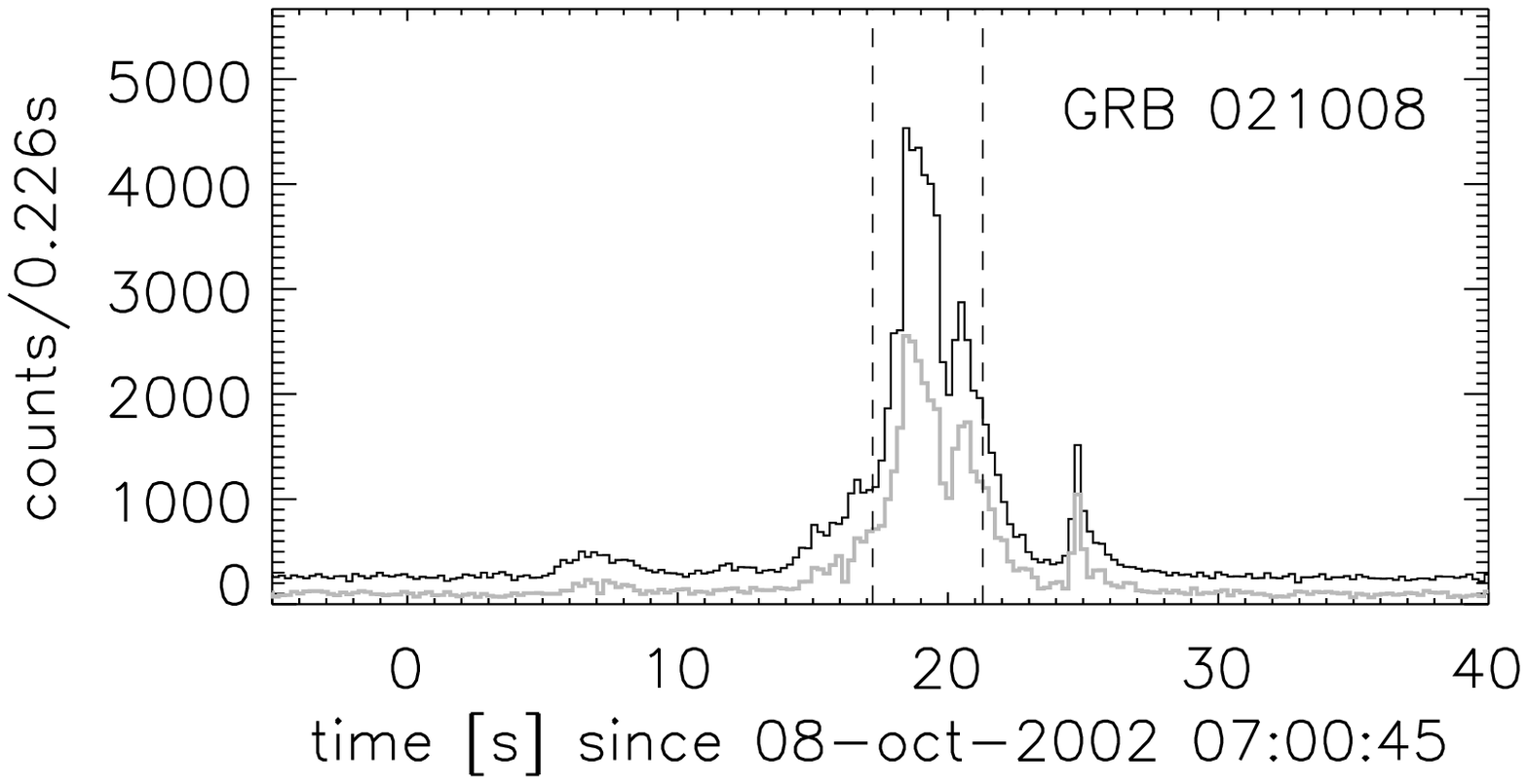} 
\plotone{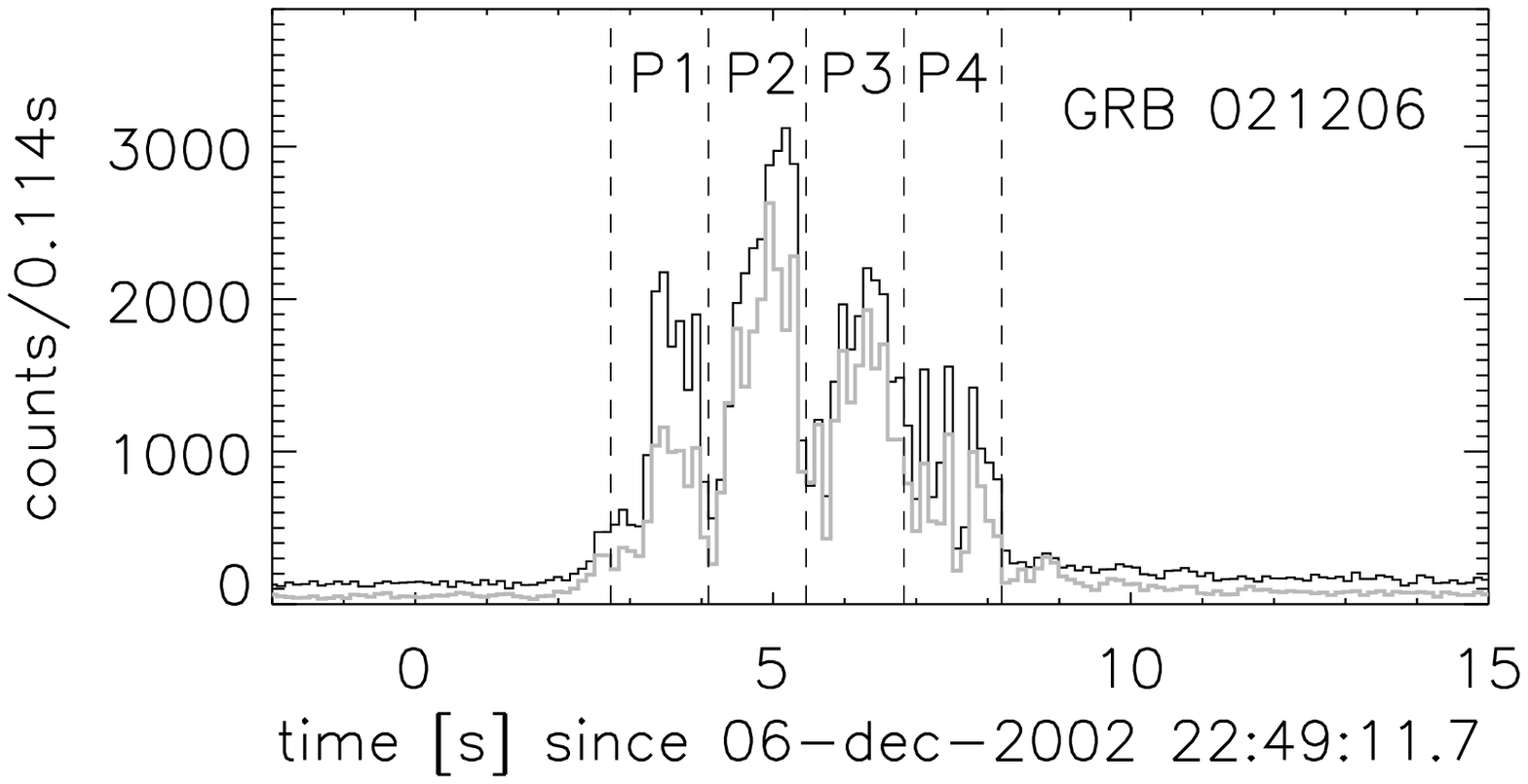} 
\plotone{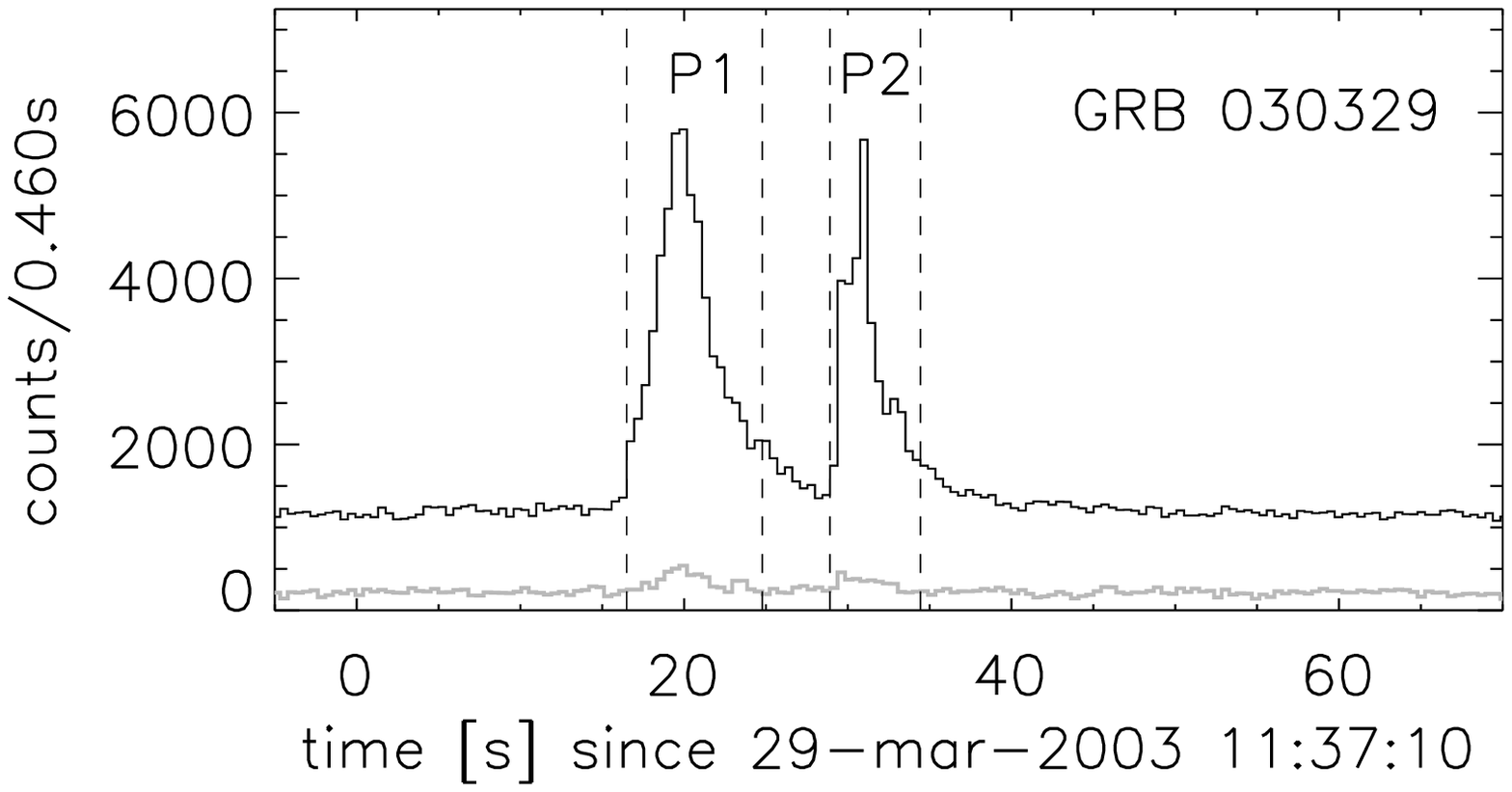} 
\caption{lightcurves
  for the energy band 20~\keV\ to 3~\MeV;
  black: rear detectors,
  grey: front detectors;
  broken vertical lines:
  time intervals used for spectral analysis.
  \label{fig:ltc_GRBsI}}
\end{figure}

\begin{figure}
\plotone{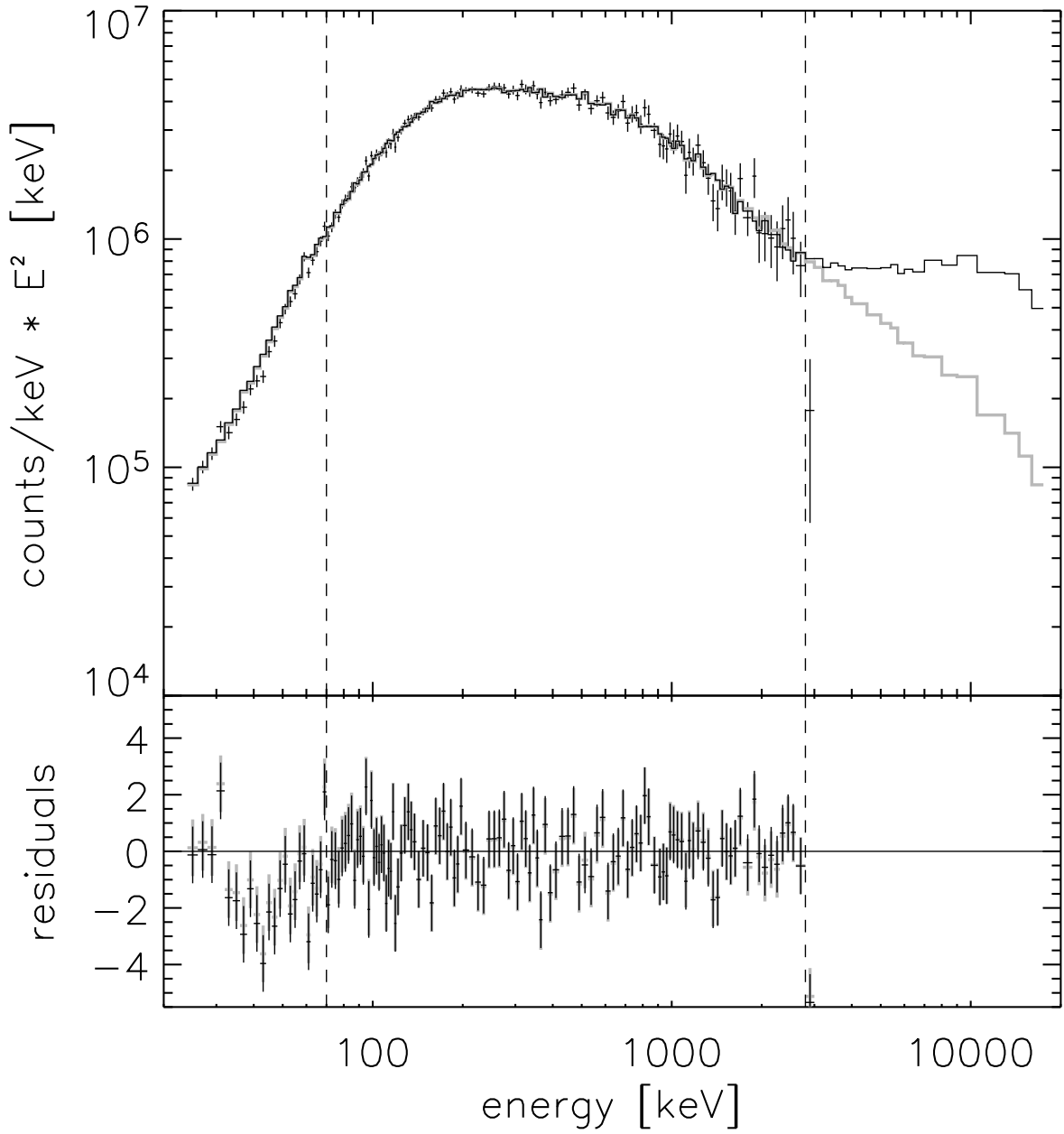} 
\caption{spectrum of GRB 021206, front detectors;
  explanations see caption of Fig.~\ref{fig:020715}.
  the same set of parameters 
  (eqs.~\ref{eq:CB021206} and \ref{eq:Band021206}) 
  is used for this plot and Fig.~\ref{fig:spec_rear}.
  \label{fig:spec_front}}
\end{figure}

\begin{figure}
\plotone{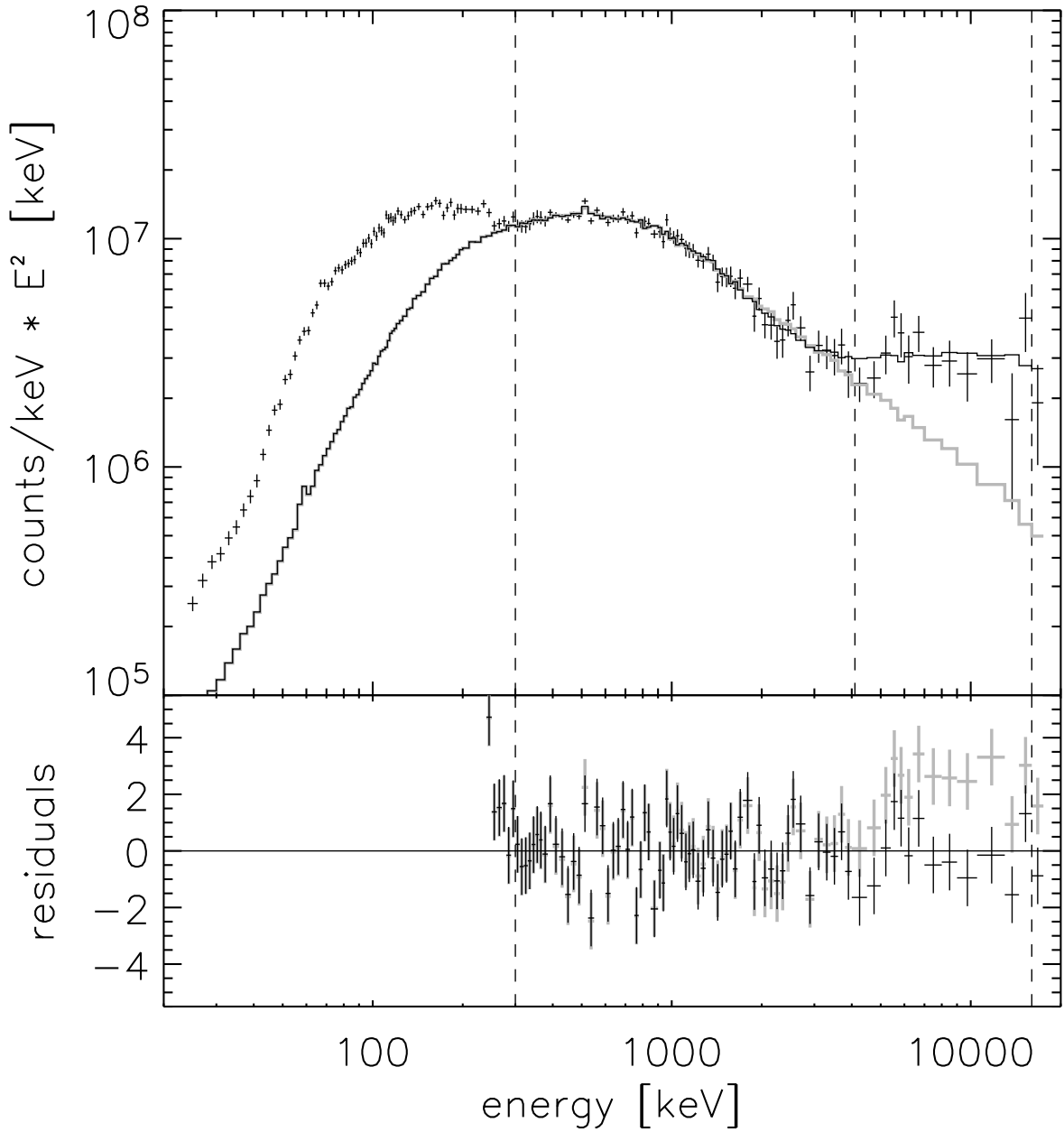} 
\caption{top: spectrum of GRB 021206, rear detectors;
   symbols as in Fig.~\ref{fig:020715}.
   bottom: residuals of \CB\ fit (black) and Band fit (grey).
   The Band function was fitted only up to 4.5~\MeV.
   The excess counts below 300~\keV\ are backscatters from Earth, see text.
   The same set of parameters 
   (eqs.~\ref{eq:CB021206} and \ref{eq:Band021206})
   is used for this plot and Fig.~\ref{fig:spec_front}.
   \label{fig:spec_rear}}
\end{figure}

This GRB is only 18 degrees from the Sun, exposing
mainly the front segments of RHESSI's detectors.
Its lightcurve is shown in Fig.~\ref{fig:ltc_GRBsI}.
Fig.~\ref{fig:spec_front} shows the energy spectrum 
in the front segments, Fig.~\ref{fig:spec_rear}
in the rear segments.

\paragraph{Fit}
The front spectrum can be fit from 70~\keV\ up to 2800~\keV,
and the rear 
from 300~\keV\ to 16~\MeV.
The huge number of excess counts below $300$~\keV\ in the 
rear detectors is understood:
The geometrical constellation of the GRB, RHESSI, and Earth
was such that the GRB photons came from the front
direction, where the effective area is relatively small,
whereas the Earth was behind RHESSI so that the
backscattered photons could easily reach the rear
segments.

The only function that fits the front {\em and} rear spectra 
over the entire energy range from 70~\keV\ up to 16~\MeV\ is the \CB.    
%
%
%
%
%
Fitting front and rear spectra simultaneously with \CB\ 
and all parameters free, yields (\CLerrors):
\begin{eqnarray}
   T_p    &= 678  \; ^{+13}_{-10}       \mbox{ \keV} \nonumber \\
   \alpha &= 0.60 \; ^{+0.09}_{-0.08}   \label{eq:CB021206}    \\
   \beta  &= 2.12 \; ^{+0.08}_{-0.13}   \nonumber              \\
    b     &= 0.108\; ^{+0.039}_{-0.045} \nonumber
\end{eqnarray}
with $\chi^2 = 187.5$ for $n_{DoF}= 185$.
These values are used for the black line histogram 
and the residuals in Figs.~\ref{fig:spec_front} and \ref{fig:spec_rear}. 

The front spectrum alone is well described
by a Band function. Its $\chi^2$ is even marginally smaller
than that of the \CB\ model fit, see Table~\ref{tab:chi2}.
The Band function also fits the rear spectrum up to 4.5~\MeV\ with 
$\chi^2 = 72.8$ (62 DoF), but not at higher energies.
The front and rear parameters (up to 4.5~\MeV) agree,
see Table~\ref{tab:Band_pars}. Evaluating the full parameter space
simultaneously for front and rear yields (\CLerrors):
\begin{eqnarray}
   E_p    &= 711  \; ^{+15}_{-17}       \mbox{ \keV} \nonumber \\
   \alpha &= 0.692\; ^{+0.033}_{-0.039} \label{eq:Band021206}  \\
   \beta  &= 3.19 \pm 0.08                           \nonumber
\end{eqnarray}
with $\chi^2=176.5$ for $n_{DoF}= 174$.
These values are used for the grey line histogram in 
Figs.~\ref{fig:spec_front} and \ref{fig:spec_rear}.
They agree with the preliminary results by \citet{cw_venice}.
Above 4.5~\MeV, a PL with $\gamma = 2.23 \pm 0.21$ fits
the data ($\chi^2 = 12.5$ for 10 DoF).

\paragraph{Discussion}
As can be learned from Table~\ref{tab:chi2},  
the high energy part can not be fit
by Band, BPL or CPL, and the 
low energy part of the spectrum can not be fit
by BBPL nor by BBmPL. 
The only function that fits over the whole RHESSI energy range
is \CB.

The \CB\ function has one parameter more than the Band function.
An F-test indicates that the chance
probability of producing such an improvement in $\chi^2$ with the
additional parameter
is only $4.0\times10^{-9}$.
The spectral hardening at 4.5~\MeV\ is significant. 

Because this GRB has so many counts at high energies,
we used a simulation with $\gamma_{sim} = 1.75$
for the results cited above.
A power law index of 1.75  results in relatively more counts
at high energies than the usual power law index (=2).
We also used simulations with $\gamma_{sim} = 1.5$
and $\gamma_{sim} = 2.0$. The results were almost
identical, especially for the high energy parameters 
$\beta$ and $b$ of the \CB\ fit.

The high energy photon index $\beta$ of the \CB\ function
agrees perfectly with the theoretical expected value ($\approx 2.1$).
The low energy photon index $\alpha \approx 0.6$ on the other hand 
is slightly smaller than expected from theory ($\approx 1.0$). 


\paragraph{Peak resolved analysis}

The time structure of GRB 021206 is rather intricate.
Four periods of emission can be distinguished, 
see Fig.~\ref{fig:ltc_GRBsI}, each of them probably 
consisting of several overlaying sub-peaks.
Luckily, these time periods match quite well our
minimum time resolution of one third of a rotation period
for fitting with a rotation averaged response function
(see \S \ref{sec:fit}).

The fitted parameters are listed in 
Table~\ref{tab:dt_grb021206} for the four time
intervals marked in the figure,
as well as the additional {\em tail} interval.
The {\em tail} interval lasts one full rotation,
starting at the end of the P4 interval.
The fluences of the two components in the \CB\ function (\refeq{eq:CB})
are listed separately (\fluenceCPL\ for the CPL component 
and \fluencemPL\ for the modified PL component).
The mPL index $\beta$ was kept frozen at 2.1.\ %

The energy $T_p$ increases from the first to the second time interval
and then decreases. Also $F_{CPL}$ and $F_{mPL}$ increase
from the first to the second interval, and then decrease.
However, the modified PL component seems to decay more slowly
than the CPL component. The tail is dominated by the mPL
component.


\subsection{GRB 030329}
GRB 030329 is famous for the supernova 2003dh detected
in its afterglow  
\citep{Hjorth2003,GCN_sn_obs,GCN2131,GCN2169}.
The authors of the \CB\ model used
the lightcurve of this GRB and its early
afterglow to predict the later afterglow and 
the appearance of a supernova
\citep[see][]{Dado_030329}.
A supernova and the late afterglow was also predicted by
\citet{Zeh2003}, and is discussed in \citet{Ferrero2006}.

In the lightcurve of GRB 030329, two peaks are
clearly separated (Fig.~\ref{fig:ltc_GRBsI}, bottom plot).
We analyse them separately.

\paragraph{Fit}
The spectrum of the first peak
is shown in Fig.~\ref{fig:030329_p1}.
Three bins around a background line from 64~\keV\ to 70~\keV\ were 
not included in the fit. 
A very good fit (see Table~\ref{tab:chi2}) is a CPL with: 
$  \mathEpeak  =  158.7 \pm 5.0 \mbox{ keV}$   
and $  \alpha    =  1.662 \pm 0.032 .$
Not surprisingly, the \CB\ and Band function, having more parameters,
but being closely related to a CPL, fit only marginally better. 

\begin{figure}
\plotone{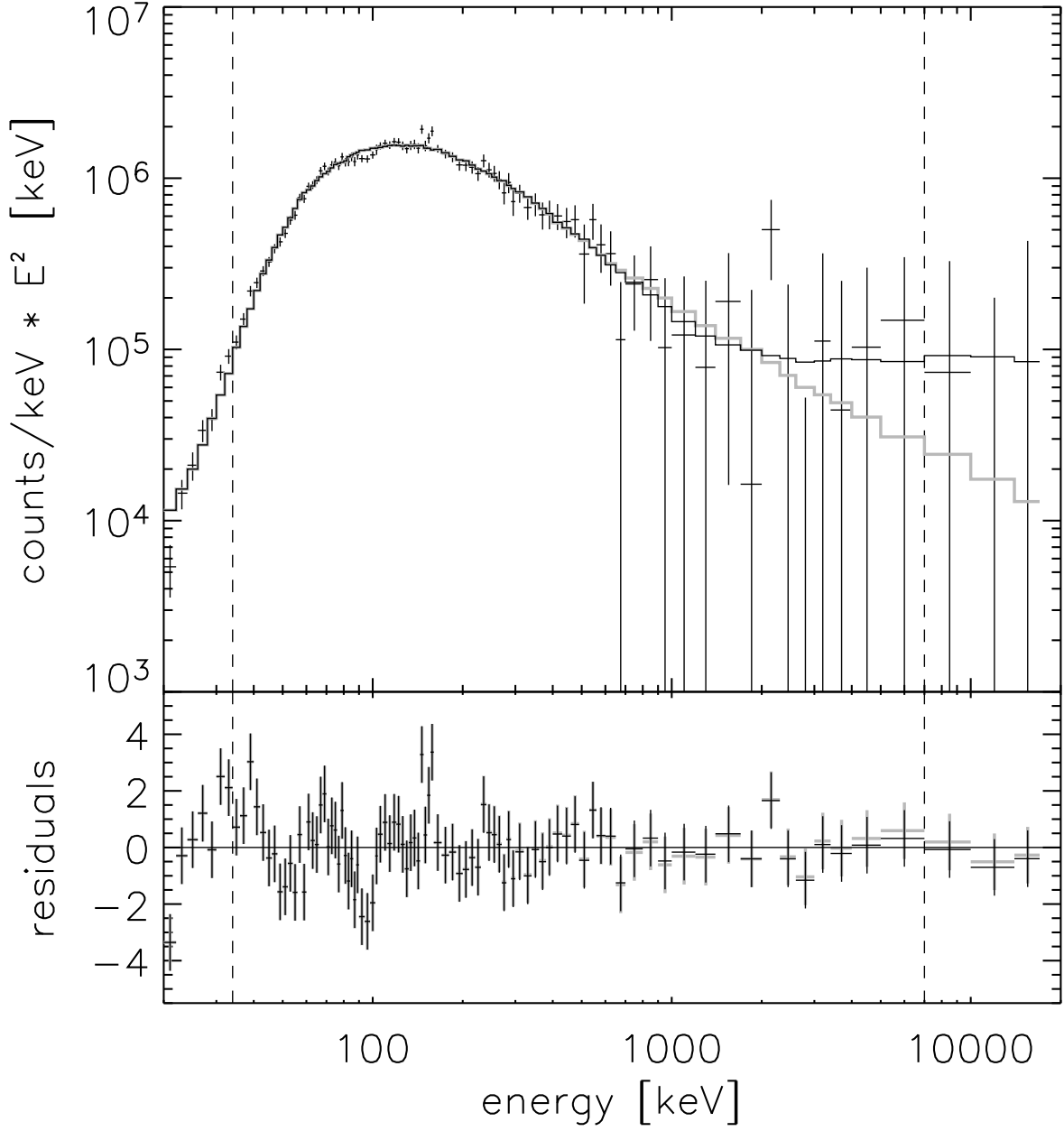} 
\caption{rear spectrum of GRB 030329, second peak;
   explanations see caption of Fig.~\ref{fig:020715}.
  \label{fig:030329_p2}}  
\end{figure}

The spectrum of the second peak
is shown in Fig.~\ref{fig:030329_p2}.
The spectrum has some wiggles below 160~\keV,
which account for the relatively high $\chi^2$.
Many models give an acceptable fit, only BBPL 
does not fit. 
A good fit (see Table~\ref{tab:chi2}) is a CPL with 
$  \mathEpeak  =  78 \pm 13 \mbox{ keV} $   
and $  \alpha    =  1.83 \pm 0.05 $.
Also a good fit is a broken power law (BPL) with 
$  E_b  =  175 \pm 15 \mbox{ keV} $,  
 $ \alpha    =  1.985 \pm 0.032$, and   
 $ \beta     =  2.68 \pm 0.08 $. 
Also Band function and \CB\ model fit the data well,
see Fig.~\ref{fig:030329_p2}.

\paragraph{Discussion}
The prompt emission was detected by HETE.
Its spectrum is published by \citet{HETE030329}
and by \citet{HETE2005}.
\citet{HETE030329} do a time resolved analysis.
For the entire burst they find (\CLerrors): 
$\mathEpeak= 70.2 \pm 2.3$~\keV,
$\alpha    = 1.32 \pm 0.02$,
$\beta     = 2.44 \pm 0.08$.
\citet{HETE2005} find for the entire burst (\CLerrors): 
$\mathEpeak= 68 \pm 2$~\keV,
$\alpha    = 1.26 \pm 0.02$,
$\beta     = 2.28 \pm 0.06$.

The RHESSI parameters, for both peaks, are all significantly higher
(see Table~\ref{tab:Band_pars}).
However, the high energy photon indices can not be compared directly,
because the break energy (above which $\beta$ is determined, 
see \refeq{eq:Band})
for HETE is 116~\keV, whereas
for RHESSI it is $>400$~\keV, 
i.e.\ above the HETE energy range.
Fitting the RHESSI data (for the entire duration of the burst)
from 135~\keV\ to 500~\keV\ only, where
the RHESSI response is good,
we find $\beta = 2.441 \pm 0.032$, in excellent agreement 
with HETE.
Fitting the RHESSI data from 400~\keV\ to 2000~\keV, i.e.\ above
the HETE range, we find $\beta = 3.11 \pm 0.25$.
The spectrum seems to soften above $\approx 350$~\keV. 

The high RHESSI value for $\alpha$ (almost 2.0) for the 
second lightcurve peak indicates
that the spectral peak (in the \EFE\ representation) is broad.
Since RHESSI's sensitivity drops below $\approx 80$~\keV, and this
is a GRB with \Epeak\ in the order of 100~\keV, 
it is likely that RHESSI's $\alpha$ describes rather the broadness
of the peak than the low energy photon index.
This opinion is supported by the fit result of the BPL fit.
The low energy photon index $\alpha \approx 2.0$ shows
that the \EFE\ spectrum is flat from 34~\keV\ to 175~\keV.


It should also be mentioned that the $\chi^2$ of the 
HETE fit, as cited by \citet{HETE2005}, is very bad.
Also the RHESSI $\chi^2$ of the spectral fits for
the second peak are rather high.
A joint fit of RHESSI and HETE data might
reveal interesting features.

\subsection{GRB 030406}

\paragraph{Fit}
The lightcurve is shown in Fig.~\ref{fig:ltc_GRBsII} (top) and 
the spectrum is shown in Fig.~\ref{fig:030406}.
The best fit is the \CB\ function with
the parameters $\alpha$ and $\beta$ frozen at 
the theoretical values.  
The next best fit is the Band function. 

\paragraph{discussion}
The spectrum of this burst was also studied by \citet{Radek2006} 
using data from
the INTEGRAL satellite. For the spectral analysis, they
used combined ISGRI and IBIS Compton mode data. 
The time interval used by \citet{Radek2006} differs from ours.
Using a similar time interval as their `peak' time interval, we find:

$\bullet$
Fit of a BPL from 24--2400~\keV\ (comparable to the energy range
in the analysis by \citet{Radek2006}): 
fits well
($\chi^2=70.3$ for $n_{DoF}=60$),
and the parameters are:
$E_b= 479 \pm 80$~\keV,
$\alpha    = 1.08 \pm 0.05$,
$\beta     = 1.96 \pm 0.02$. 

$\bullet$
Fit of a BPL from 24~\keV\ to 16~\MeV: does not fit well
 ($\chi^2 = 86.0$ for $n_{DoF}=71$).

$\bullet$
Fit of \CB\ from 24~\keV\ to 16~\MeV\ with $\alpha=1.0$ and 
$\beta=2.1$ frozen: fits well
($\chi^2 = 77.9$ for $n_{DoF}=72$), and the parameters are:
$T= 1220 \pm 110$~\keV\ and $b = 0.055 \pm 0.079$,

$\bullet$
Fit of Band function: fits well ($\chi^2 = 77.33$ for $n_{DoF}=71$), 
and the parameters are: 
$\mathEpeak= 1180 \pm 120$~\keV,
$\alpha    = 0.96 \pm 0.06$,
$\beta     = 3.02 \pm 0.55$. 

For the high energy part, 
the parameters found by \citet{Radek2006}  
and by us agree. 
But we can not confirm their hard low energy photon index $\alpha < 0.0$.
We even dare to say that we trust our low energy photon index better,
because for this GRB incoming direction, the RHESSI response function
is well understood, whereas the INTEGRAL response function of this burst
might suffer from same systematic effects that we described for RHESSI in \S \ref{sec:systematics}.

\subsection{GRB 030519B}
\paragraph{Fit}
The lightcurve is shown in 
Fig.~\ref{fig:ltc_GRBsII}
and the spectrum in Fig.~\ref{fig:030519B}.
%
The best fit is the Band function,
followed by \CB.

\paragraph{Discussion}
In the HETE GRB catalog by \citet{HETE2005} one finds
(\CLerrors):
$E_{p}= 138\; ^{+18}_{-15} $~\keV,
$\alpha= 0.8 \pm 0.1$,
$\beta = 1.7 \pm 0.2$.
Since $\beta < 2.0$, the energy $E_p$ is not the 
peak energy, but only a variable related to the parameter
$E_0=E_p/(2-\alpha)$.
Indeed, the RHESSI peak energy for this GRB is $>400$~\keV.
But the HETE parameters do not fit the RHESSI
spectrum from 70~\keV\ to 350~\keV\ ($\chi^2 = 141$ for 
53 energy bins).


\subsection{GRB 031027}
\begin{figure}
\plotone{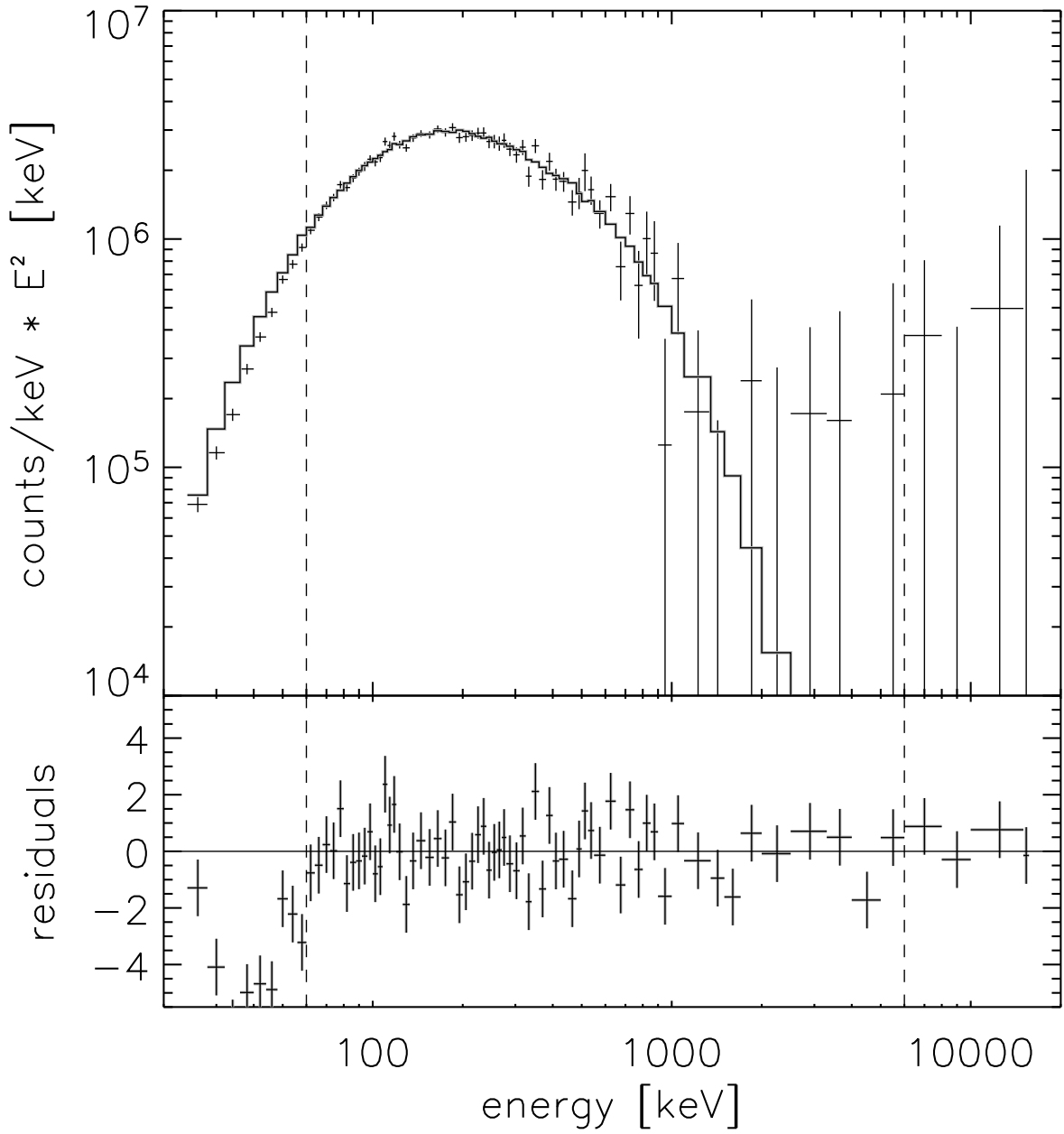} 
\caption{rear spectrum of GRB 031027;
  histogram and residuals: fit of a cut off power law (CPL);
  a CPL is equivalent to a Band function with
  $\beta = \infty$  or \CB\ with $b=0$.
  \label{fig:031027}}  
\end{figure}

The lightcurve  is shown in 
Fig.~\ref{fig:ltc_GRBsII} and 
the spectrum in Fig.~\ref{fig:031027}.
The best fit is a cut off power law with 
$ \mathEpeak = 336 \pm 9$~\keV\ and $\alpha = 0.940\pm 0.05$.
This function is shown in Fig.\  \ref{fig:031027}.
Band function and \CB\ have difficulties to converge.
Since the CPL fits so well, we expect $\beta = \infty$ 
for the Band function and $b=0$ for the \CB\ function.

\subsection{GRB 031111}
\paragraph{Fit}
The lightcurve is shown in Fig.~\ref{fig:ltc_GRBsII} (bottom) and 
the spectrum is shown in Fig.~\ref{fig:031111}.
Band function, \CB\ and BBmPL are good fits, 
the best being Band function.

\paragraph{Discussion}
A preliminary CPL fit to the HETE data is
published on a web page\footnote{
\url{http://space.mit.edu/HETE/Bursts/GRB031111A/}
}
as $E_{0}= 600.5$~\keV\ and 
$\alpha= 0.8366$ with a good $\chi^2$.
These values describe the RHESSI spectrum
well from 80~\keV\ to 350~\keV, but not at
higher energies.
HETE's energy range ends at 400~\keV, 
thus we believe that our values for $E_0$ and $\alpha$ are better.

\section{GENERAL DISCUSSION AND CONCLUSION}
\label{sec:discussion}

\subsection{The spectral functions}

What is an acceptable $\chi^2$? 
In the limit of many degrees of freedom 
($n_{DoF} > 30$), $\chi^2$ is normal distributed
with an expectation value of $n_{DoF}-0.5$ and
a variance of $\sigma_{\chi^2} = \sqrt{2n_{DoF}-1}$.
A fit is acceptable if $\chi^2$ is close to its expectation
value {\em and} if the residuals
scatter around zero over the whole fit range, 
i.e.\ if the fit ``looks good''.

From Table~\ref{tab:chi2} we conclude that
the \CB\ gives acceptable $\chi^2$ for {\em all} GRBs studied.
And they also look good, as can be seen in 
Figs.~\ref{fig:020715} to \ref{fig:031111}.
Except for GRB 021206, rear (Fig.~\ref{fig:spec_rear}), 
the same can be said for the Band function.

In many cases, a cut off power law (CPL) 
fits the spectrum up to
high energies, e.g.\ GRB 021008, GRB 030329 or GRB 031027.
In these cases, Band and \CB\ 
improve the goodness-of-fit slightly, but all three spectral
shapes fit the data acceptably.

A broken power law fits sometimes, but usually not well. 

BBPL and BBmPL do not fit in general, BBPL worse than BBmPL.
However, it should be mentioned that a blackbody component
is expected---if at all---only
at the beginning of a GRB (see e.g.\ \citet{Ryde2006}
and references therein),
whereas we fitted the entire duration of the bursts.
When using BBPL, we often find that the PL component
fits either at high energies or at low energies.
This is also discussed by \citet{Ghirlanda2007}
who studied six BATSE GRBs in detail,
where low energy data from the WFC instrument (on board BeppoSAX) are
available. They find that the WFC data fit
the Band function or CPL extrapolation, but not the
BBPL extrapolation to low energies. 
Arguing that the PL contribution is too simple, they
try to fit a blackbody spectrum plus CPL.  
We suggest to use our BBmPL function instead.
Its modified PL component describes a spectral break from
$dN/dE \propto E^{-(\beta-1)}$ at low energies to
$dN/dE \propto E^{-\beta}$ at high energies.

\subsection{\CB\ function }
 
The present work is, to our knowledge, the first systematic 
attempt to fit the \CB\ function to prompt GRB spectra.
The two terms in \refeq{eq:CB}
have a simple meaning.
According to the cannonball theory, all GRBs are associated with a supernova.
The ambient light is Compton up-scattered by the cannonball's electrons,
producing the prompt GRB emission.
Some electrons are simply comoving with the cannonball, 
giving rise to the CPL term in \refeq{eq:CB}.
Since the photon spectrum of the ambient light can be described
by a thin thermal bremsstrahlung spectrum,
$\alpha$ is expected to be $\approx 1$.
The second term (mPL) is caused by
a small fraction of electrons accelerated 
to a power law distribution, resulting in a 
photon index of $\beta \approx 2.1$.
See e.g.\ \citet{Dado2005}, \S 3.8. 
or \citet{Dado2007}, \S 2 and 4.1 for a summary. 

In our study, the observed values for $\alpha$ are all approximately 1,
as predicted by the \CB.
Because of the low count statistics at high energies, 
we could not always fit $\beta$.
We then fixed it to its theoretical value
of 2.1 in order to make the fit converge and to obtain
a value for the parameter $b$.
In the cases where we could fit $\beta$,
we found values close to 2.1 (Table~\ref{tab:CB_pars}). 

For the factor $b$ of the modified PL component in the \CB\ function
we typically found values of the order $0.1\,$.
An exception is GRB 031111, where $b$ is of
the order 1.0, but with a large error ($0.4$).

Our values for $\beta$ and $b$ are similar to the 
ones found by 
\citet{Dado2005} (fit of GRB 941017)
and by
\citet{Dado2004} (fit  of $X$-ray flashes XRF 971019, XRF 980128,
and XRF 990520 using BeppoSAX/WFC and CGRO/BATSE data).
The authors of the \CB\ hypothesize that XRFs are simply GRBs viewed
further off the jet axis.

The different time development of the CPL- and the mPL-fluences,
as reported in Table~\ref{tab:dt_grb021206}, 
possibly point to a different time dependence of 
the two underlying electron distributions within a cannonball.

\subsection{Fitting \CB\ function versus CPL and Band function}
Both the Band function and the \CB\ function are extensions
of a CPL, the Band function with one additional parameter,
the \CB\ function with two. For cases where a CPL fits the
data well, also a Band function with $\beta = \infty$ or
a \CB\ function with $b=0$ (and $\beta=2.1$ or any other value) 
fits. This is the case for GRB 031027.

Whether additional parameters are necessary in a fit,
can be tested with the $F$-test.
For GRB 030329, the extra parameters are barely needed.
For GRB 020715, 021008, 021206, 030406, 030519B, 031111
additional parameters are required at a confidence level
of at least 90\%.

Concerning the question of whether the high energy power law
parameter $\beta$ in the \CB\ should be treated as free parameter,
the answer is 'yes' from a theoretical point of view, but
in practice, see Table~\ref{tab:chi2}, the improvements in $\chi^2$ 
are marginal or small for all bursts we studied.
Our practice to freeze $\beta$ at its theoretically predicted
value in cases of bad convergence
seems to be acceptable.
 
It is more difficult to compare the goodness of fit using
the Band function compared to using the \CB\ function.
The two functions are not independent, because they both
are dominated by a CPL up to the peak energy and higher.
In most cases of our study, the two functions fit the observed
spectrum equally well with a slight preference
for the Band function. 
At high energies however (typically above several
times the peak energy) the two functions are different, the spectral
hardening being a unique feature of the \CB\ function.
There is only one case, namely GRB 021206, where this
hardening is observed. For the rear data going up to high energies, 
a Band function fit gives $\chi^2 / \mbox{dof}
= 133.3 / 74$ (see Table~\ref{tab:chi2}). This is not 
acceptable at $<0.01$\% probability of being accidentally so high.
The \CB\ fit on the other hand gives  $\chi^2 / \mbox{dof}
= 82.7 / 73$, which is fully acceptable at a $20$\% level.

We would like to stress again that, while the \CB\ gives acceptable
fits for {\em all} cases, the Band function fails in one case.
This seems enough to us to give some credit to the \CB.

But it is, of course, no proof that the \CB\ is the only theory capable
of describing the spectrum of GRB 021206. 
For example, a Band function plus a PL with $\gamma\approx 1.5$ 
would also fit.
But there is no theory to predict such a shape.
To our knowledge, \CB\ is to date the only existing GRB model 
that explains the prompt GRB spectra from first principles. 

At this place we also would like to note the the mean 
$\alpha$-value found for the BATSE catalogue is 1 
\citep[see][]{BATSEcatalog}). We cite from their summary:
``{\em We confirmed, using a much larger sample, that
the most common value for the low-energy index is $\approx -1\,$}\footnote{ 
this corresponds to $+1$ in our notation}
{\em \citep{Preece2000, Ghirlanda02}. 
 The overall distribution of this parameter 
shows no clustering or distinct features at the values expected from
various emission models, such as $-2/3$ for synchrotron 
\citep{Katz94, Tavani96},
$0$ for jitter radiation \citep{Medv00}, 
or $-3/2$ for cooling synchrotron \citep{GhisCel99}.}''
They do not mention the \CB\ which would explain $\alpha \approx 1$.

Note that the $\beta$ values of the \CB\ are systematically lower 
than the $\beta$ values of the Band function,
compare Tables \ref{tab:CB_pars} and \ref{tab:Band_pars}.
From Band function fits to BATSE GRBs, it is known that
$\beta$ is clustered around 2.3, with a long tail
towards higher values,   see \citet{BATSEcatalog}.
For \CB\ we would expect $\beta$ to cluster at
slightly lower values.

For criticism of the \CB, see e.g.\ \citet{Hillas}, but see
also the answer by \citet{Dar2006}.

\subsection{The spectral hardening}

The difference of a \CB\ spectrum and
Band function is the hardening at high energies. 
This becomes visible---for the GRBs studied here---in 
the few \MeV\ region, but it depends on the peak energy
and the factor $b$. For $\alpha=1.0$ and $b=0.10$ the hardening
typically appears at several times the peak energy
and the second term dominates at 10 times the peak energy.
For the spectral fit of XRFs done by \citet{Dado2004}, 
the spectral hardening is expected in the few hundred \keV\ region,
just where the number of photons detected runs low.
Most of our GRBs also suffer from this lack of statistics at high energies,
preventing the detection of a hardening.

A spectral coverage of two decades and 
good detection efficiency at high energies
is necessary to experimentally observe the 
full shape of the \CB\ function. 
In the case of GRB 021206 we were able to detect this hardening, 
thanks to RHESSI's broad energy range (30~\keV\ to 15~\MeV),
and because this is one of the brightest GRBs ever observed.

There is a GRB observed by SMM from 20 keV up to 100 MeV,
namely GRB 840805.
As reported by \citet{Share84},
the spectrum of this burst shows emission up 100 MeV. 
In order to fit the spectrum, ``a classical thermal synchrotron function
plus a power law'' was used. The power law component was 
required to fit the data above about 6 MeV. This is a hint
of a spectral hardening around 6 MeV, and we suppose that
the spectrum of this GRB can be fit by a \CB\ function.
 
The spectral hardening observed in GRB 941017
\citep{Nature2003} seems to be different. 
The photon index of GRB 941017 above a few \MeV\ is $\approx 1.0$.
This case is discussed by \citet{Dado2005} as a possible 
additional feature in the \CB\ spectrum.

\subsection{Outlook}
\label{sec:outlook}

In order to find more GRB spectra that show the hardening
characteristic for the \CB\ function,
strong GRBs have to be observed over a broad enough energy range.
With the forthcoming GLAST mission, we expect 
that more such spectra will be observed.
But also joint analyses with more than one instrument
could reveal this hardening.
We therefore suggest:


$\bullet$ 
to search for \CB\ spectrum candidates among joint Swift/RHESSI
GRBs and XRFs, and joint Swift/Konus GRBs.

$\bullet$ 
to reanalyse some BATSE bursts. Looking
at the BATSE spectra published by \citet{Ghirlanda2007}, 
we suppose that the \CB\ can possibly improve the fits of 
GRB 980329, GRB 990123, and GRB 990510. 
The same can be said for  
GRB 911031 as published by \citet{Ryde2006}. 
And GRB 000429, as published in Fig.\ 19 of \citet{BATSEcatalog},
looks like a promising candidate as well.

$\bullet$ 
to search in KONUS data for suitable GRBs.

$\bullet$ 
to add the \CB\ function to XSPEC
in order to make it more accessible
to the astronomical community.

\section{SUMMARY}
\label{sec:summary}

We have presented the time integrated spectra of 8 bright
GRBs observed by RHESSI in the years 2002 and 2003.

The spectrum of GRB 021206 shows a hardening above 4~\MeV. 
From 70~\keV\ to 4.5~\MeV, the spectrum can be well
fitted by a Band function -- but not above that.
The cannonball model successfully describes
the entire spectrum up to 16~\MeV, the upper limit
of RHESSI's energy range.
For the spectra of the seven other GRBs analysed, we found that they
can be fitted by the \CB\ as well as by the Band function.

We therefore suggest that the cannonball model should
be considered for fitting GRB spectra.



\acknowledgments

We thank K. Hurley, A. Kann, S. McGlynn, J. \v{R}ipa and
L.\ Hanlon for helpful discussion and comments.






\clearpage
\begin{deluxetable}{lcccccc}
\tablecolumns{7}
\tablecaption{GRB analysis time intervals
         \label{tab:GRBs}}
\tablewidth{0pt}
\tablehead{
\colhead{GRB} & 
\colhead{$t_0$} & 
\colhead{$\Delta t_{burst}$} &
\colhead{$\Delta t_{BG1}$} & 
\colhead{$\Delta t_{BG2}$} & 
\colhead{$\theta$} & 
\colhead{ref.}
\\
\colhead{} &
\colhead{(UT)} & 
\colhead{(s)} &
\colhead{(s)} &
\colhead{(s)} &
\colhead{(degrees)} & 
\colhead{}
}
\startdata
020715    & 19:20:56.0 & [11.53,15.55]       & [\phn-80.46,0.0]  & [\phn48.28,168.97]    &\phn72.4 & 1 \\
021008    & 07:00:45.0 & [17.21,21.29]       & [\phn-73.37,0.0]  & [\phn36.68,\phn48.91] &\phn50.1 & 2 \\
021206    & 22:49:11.7 & [\phn2.73,\phn8.19] & [\phn-53.26,0.0]  & [\phn20.49,102.43]    &\phn18.0 & 3 \\
030329 P1 & 11:37:10.0 & [16.56,24.84]       & [\phn-70.39,0.0]  & [\phn70.39,140.78]    &   144.1 & 4 \\  
030329 P2 &    "       & [28.98,34.50]       &       "           &            "          &     "   & " \\  
030406    & 22:41:30.0 & [85.68,89.83]       & [-140.96,0.0]     & [140.96,281.93]       &\phn96.1 & 5 \\  
030519B   & 14:04:53.0 & [\phn0.46,11.47]    & [\phn-61.94,0.0]  & [\phn28.90,\phn90.84] &   165.5 & 6 \\ 
031027    & 17:07:06.0 & [29.71,45.92]       & [-137.77,0.0]     & [\phn68.88,206.65]    &   101.5 & 7 \\
031111    & 16:45:12.0 & [\phn2.27,\phn6.35] & [-122.51,0.0]     & [\phn12.25,134.76]    &   155.6 & 8 \\  
\enddata                                                                                          
\tablecomments{$t_0$: reference time; 
  $\Delta t_{burst}$: time interval for spectral analysis; 
  $\Delta t_{BG1}$: background time interval before GRB; 
  $\Delta t_{BG2}$: background time interval after GRB;
  time intervals are given relative to $t_0$.
  $\theta$: angle between GRB direction and RHESSI axis;
  References: 
  (1) GCN 1456, 1454 \citep{GCN_020715_IPN},	 
  (2) GCN 1629, 1617 \citep{GCN_021008_IPN},		
  (3) GCN 1728, 1727 \citep{GCN_021206_IPN},	
  (4) GCN 1997 \citep{GCN_030329_HETE},		
  (5) GCN 2127 \citep{GCN_030406_IPN},
  (6) GCN 2235, 2237 \citep{GCN_030519B_HETE, GCN_030519B_IPN},	
  (7) GCN 2438 \citep{GCN_031027_IPN},		
  (8) HETE trigger 2924, GCN 2443 \citep{GCN_031111_IPN}.}
\end{deluxetable}

\begin{deluxetable}{lcccrrrrrrr}
\tabletypesize{\small}
\tablecolumns{11}
\tablecaption{Chi-square of spectral fits
	 \label{tab:chi2}}
\tablewidth{0pt}
\tablehead{
\colhead{GRB} & 
\colhead{$\Delta E_{\mbox{\scriptsize front}}\,$} & 
\colhead{$\Delta E_{\mbox{\scriptsize rear}}\,$} & 
\colhead{$n$} & 
\colhead{CPL} & 
\colhead{Band} &
\colhead{\CB} &
\colhead{\CB} &
\colhead{BPL} &
\colhead{BBPL} &
\colhead{BBmPL}
\\ 
\colhead{} &
\colhead{(\keV)} &
\colhead{(\keV)} &
\colhead{} &
\colhead{$n_{par}=3$} &
\colhead{4} &
\colhead{4} &
\colhead{5} &
\colhead{4} &
\colhead{4} &
\colhead{4}
}
\startdata
020715     & \nodata       &[\phn30,15660]   & 117 &  113.9  & 106.3   & 110.8   & 110.7   &  129.2  &  270.0  & 157.1   \\ \hline
021008     &[300,2800]     &\nodata          &  38 &   35.5  &  34.8   &  35.1   &  35.0   &   33.0  &   32.9  &  33.9   \\ 
021008     &\nodata        &[300,15660]      &  50 &   43.8  &  39.2   &  39.9   &  39.2   &   52.9  &   97.4  &  60.2   \\
021008     &[300,2800]     &[300,15660]      &  88 & \nodata &  79.5   &  80.8   & \nodata & \nodata & \nodata & \nodata \\ \hline 
021206     &[\phn70,2800]  &\nodata          & 112 &  130.6  & 103.6   & 104.1   & 103.9   &  155.5  &  315.7  & 191.8   \\  
021206     &\nodata        &[300,16000]      &  78 &  338.1  & 133.3   &  82.7   &  82.7   &  132.9  &   94.1  & 110.5   \\
021206     &[\phn70,2800]  &[300,16000]      & 190 & \nodata & \nodata & \nodata & 187.5   & \nodata & \nodata & \nodata \\
021206     &\nodata        &[300,\phn4500]   &  66 & \nodata &  72.8   &  69.8   & \nodata & \nodata & \nodata & \nodata \\
021206     &[\phn70,2800]  &[300,\phn4500]   & 178 & \nodata & 176.5   & \nodata & \nodata & \nodata & \nodata & \nodata \\ \hline
030329 P1  &\nodata        &[\phn34,10000]   &  94 &   86.5  &  84.3   &  84.8   &  84.7    &  89.3 & 137.8 &  89.1 \\ 
030329 P2  &\nodata        &[\phn34,\phn7000]&  87 &  104.5  & 103.3   & 103.2   & 103.1    &  98.9 & 102.0 &  98.4 \\ \hline
030406     &\nodata        &[\phn24,15000]   &  75 &   79.1  &  75.6   &  75.0   &  75.0    &  88.6 & 151.2 &  88.7 \\ \hline
030519B    &\nodata        &[\phn70,15000]   &  79 &  102.2  &  86.3   &  91.1   &  89.0    &  99.8 & 189.8 & 109.8 \\ \hline 
031027     &\nodata        &[\phn60,\phn6000]&  63 &   63.6  &\phn n.c.&\phn n.c.& \phn n.c.&  93.0 & 138.9 &  99.6 \\ \hline 
031111     &\nodata        &[\phn38,15000]   & 117 &  182.4  & 128.3   & 133.2   & 130.4    & 140.5 & 266.0 & 133.0
\enddata     
\tablecomments{$\chi^2$ obtained by fitting different spectral models to the data.
 $\Delta E_{\mbox{\scriptsize front/rear}}$: energy interval used to fit front/rear detector data;
 $n$: number of energy bins;
 $n_{par}$: number of free fit parameters;
 CPL: cut off power law \refeq{eq:CPL},
 Band: Band function \refeq{eq:Band},
 \CB: cannonball model \refeq{eq:CB},
 BPL: broken power law \refeq{eq:BPL},
 BBPL: blackbody plus power law \refeq{eq:BBPL},
 BBmPL: blackbody plus modified power law \refeq{eq:BBmPL};
 n.c.: fit did not converge. In the case of \CB\ with 4 parameters, $\beta$ was fixed to its theoretically
 expected value of $2.1 \,$.
 For each fit, the degree of freedom is $n_{DoF}=n-n_{par}$.
}
\end{deluxetable}

\begin{deluxetable}{lcccccc}
\tablecolumns{7}
\tablecaption{Fit results for \CB
\label{tab:CB_pars}}
\tablewidth{0pt}
\tablehead{ 
\colhead{GRB} & 
\colhead{$\Delta E_{\mbox{\scriptsize front}}\,$} & 
\colhead{$\Delta E_{\mbox{\scriptsize rear}}\,$} & 
\colhead{$T_{p}\,$} & 
\colhead{$\alpha$} & 
\colhead{$\beta$} & 
\colhead{$b$} 
\\ 
\colhead{} &
\colhead{(\keV)} &
\colhead{(\keV)} &
\colhead{(\keV)} &
\colhead{} &
\colhead{} &
\colhead{}
}
\startdata
020715    &\nodata	  &[\phn30,15660]   & 532\PM20    & $0.741 \pm 0.077$     & 2.20\PM0.14    & \phS0.067\PM0.040       \\ \hline
021008    &[300,2800]	  &\nodata	    & 628\PM71    & $1.31 \pm 0.28$       & 2.1            & \phS0.052\PM0.076       \\ 
021008    &\nodata	  &[300,15660]      & 672\PM68    & $1.487 \pm 0.062$     & 2.77\PM0.55    & \phS0.085\PM0.092       \\
021008    &[300,2800]	  &[300,15660]      & 641\PM32    & $1.523 \pm 0.055$     & 2.1            & \phS0.020\PM0.008       \\ \hline
021206    &[\phn70,2800]  &\nodata          & 672\PM20    & $0.66 \pm 0.21$       & 1.92\PM0.67    & \phS0.063\PM0.142       \\ 
021206    &\nodata        &[300,16000]      & 672\PM24    & $0.67 \pm 0.19$       & 2.12\PM0.13    & \phS0.102\PM0.048       \\
021206    &[\phn70,2800]  &[300,16000]      & 678\PM\phn6 & $0.60 \pm 0.06$       & 2.10\PM0.08    & \phS0.103\PM0.028       \\
021206    &\nodata        &[300,\phn4500]   & 670\PM23    & $0.71 \pm 0.15$       & 2.1            & \phS0.091\PM0.012       \\
021206    &[\phn70,2800]  &[300,\phn4500]   & \nodata     & \nodata               & \nodata        & \nodata                 \\ \hline
030329 P1 &\nodata        &[\phn34,10000]   & 147\PM10    & $1.614 \pm 0.036$     & 2.1            & \phS0.033\PM0.029       \\
030329 P2 &\nodata        &[\phn34,\phn7000]& \phn69\PM15 & $1.841 \pm 0.049$     & 2.1            & \phS0.048\PM0.055       \\ \hline
030406    &\nodata	  &[\phn24,15000]   & 626\PM83    & $0.966 \pm 0.089$     & 2.1            & \phS0.18\phn\PM0.12\phn \\ \hline 
030519B   &\nodata	  &[\phn70,15000]   & 396\PM12    & $0.949 \pm 0.073$     & 2.388\PM0.097  & \phS0.135\PM0.048       \\ \hline
031027    &\nodata	  &[\phn60,\phn6000]& 340\PM17    & $0.950 \pm 0.055$     & 2.1            &    -0.010\PM0.025       \\ \hline
031111    &\nodata	  &[\phn38,15000]   & 690\PM45    & $0.68 \pm 0.27$       & 2.241\PM0.023  & \phS1.09\phn\PM0.36\phn
\enddata                                                                                                                      
\tablecomments{
  $\Delta E_{\mbox{\scriptsize front/rear}}$: energy interval used to fit front/rear detector data;
  $T_{p}, \alpha, \beta, b$: parameters as defined in the text below \refeq{eq:CB};
  errors are symmetric $1 \sigma$ errors;
  where no error is given, the parameter was frozen at that value.
  }
\end{deluxetable}

\begin{deluxetable}{lccccc}
\tablecolumns{6}
\tablecaption{Fit results for Band function
         \label{tab:Band_pars}}
\tablewidth{0pt}
\tablehead{
\colhead{GRB} & 
\colhead{$\Delta E_{\mbox{\scriptsize front}}$} & 
\colhead{$\Delta E_{\mbox{\scriptsize rear}}$} & 
\colhead{$E_{p}\,$}   & 
\colhead{$\alpha$}   & 
\colhead{$\beta$}
\\ 
\colhead{} &
\colhead{(\keV)} &
\colhead{(\keV)} &
\colhead{(\keV)} &
\colhead{} &
\colhead{}
}
\startdata
020715     & \nodata      &[\phn30,15660]    & $531 \pm 24$    & $0.776 \pm 0.044$ & $3.14 \pm 0.25$   \\ \hline
021008     &[300,2800]    & \nodata          & $670 \pm 58$    & $1.36 \pm 0.23$   & $3.41 \pm 0.51$   \\ 
021008     & \nodata      &[300,15660]       & $678 \pm 42$    & $1.526 \pm 0.067$ & $3.86 \pm 0.25$   \\
021008     &[300,2800]    &[300,15660]       & $677 \pm 33$    & $1.493 \pm 0.056$ & $3.73 \pm 0.18$   \\ \hline
021206     &[\phn70,2800] & \nodata          & $713 \pm 17$    & $0.694 \pm 0.031$ & $3.20 \pm 0.13$   \\
021206     & \nodata      &[300,16000]       & \nodata         & \nodata           & \nodata           \\
021206     &[\phn70,2800] &[300,16000]       & \nodata         & \nodata           & \nodata           \\
021206     & \nodata      &[300,\phn4500]    & $709 \pm 18$    & $0.72 \pm 0.20$   & $3.186 \pm 0.063$ \\
021206     &[\phn70,2800] &[300,\phn4500]    & $711 \pm 7\phn$ & $0.692 \pm 0.020$ & $3.19 \pm 0.04$   \\ \hline
030329 P1  & \nodata      &[\phn34,10000]    & $157.2 \pm 5.2$ & $1.608 \pm 0.038$ & $3.48 \pm 0.53$   \\  
030329 P2  & \nodata      &[\phn34,\phn7000] & $\phn85 \pm 11$ & $ 1.781\pm 0.065$ & $3.04 \pm 0.30$   \\ \hline  
030406     & \nodata      &[\phn24,15000]    & $674 \pm 70$    & $0.979 \pm 0.064$ & $2.61 \pm 0.27$   \\ \hline
030519B    & \nodata      &[\phn70,15000]    & $417 \pm 13$    & $1.048 \pm 0.042$ & $3.11 \pm 0.18$   \\ \hline
031027     & \nodata      &[\phn60,\phn6000] & $338 \pm 15$    & $0.940 \pm 0.079$ & \nodata           \\ \hline
031111     & \nodata      &[\phn38,15000]    & $844 \pm 59$    & $1.102 \pm 0.036$ & $2.364 \pm 0.068$ 
\enddata     
\tablecomments{
   $\Delta E_{\mbox{\scriptsize front/rear}}$: energy interval used to fit front/rear detector data;
   $E_{p}, \alpha, \beta$: parameters as defined in the text below \refeq{eq:Band};
   where no $\beta$ is given, a CPL (\refeq{eq:CPL}) was fitted;
   errors are symmetric $1\sigma$ errors. 
  } 
\end{deluxetable}

\begin{deluxetable}{lrrrrccc}
\tablecolumns{8}
\tablecaption{Fluences in $10^{-5}\,$erg cm$^{-2}$
	 \label{tab:fl}}
\tablewidth{0pt}
\tablehead{
\colhead{GRB} & 
\colhead{\fluenceCB} & 
\colhead{\fluenceBand} & 
\colhead{\fluenceRHESSI}   &
\colhead{\fluenceHETE}   &
\colhead{HETE}  &
\colhead{\fluenceUlysses}    &
\colhead{Ulysses}   
\\
\colhead{} &
\multicolumn{7}{c}{($10^{-5}\,$erg cm$^{-2}$)}
}
\startdata
 020715       &  4.37 &  4.41 &  3.94 &  1.93    & \nodata           &  0.43    &  0.30   \\  \hline
 021008 front & 26.85 & 26.85 & 41.24 & 22.92\fa & \nodata           &  8.67\fa & \nodata \\ 
 021008 rear  & 35.54 & 35.64 & 48.85 & 27.15\fa & \nodata           & 10.27\fa &   8.5   \\  \hline 
 021206 front & 52.15 & 52.44 & 55.67 & 19.97    & \nodata           &  3.81    & \nodata \\ 
 021206 rear  & 58.74 & 53.45 & 70.55 & 25.29    & \nodata           &  4.82    &  16     \\  \hline
 030329 P1    &  6.51 &  6.47 &  4.42 &  5.20    & \nodata           &  2.57    & \nodata \\  
 030329 P2    &  3.58 &  3.58 &  2.26 &  2.95    & \nodata           &  1.69    & \nodata \\  
 030329 total &\nodata&\nodata&  7.35 &  9.46    & $10.76 \pm 0.14$  &  4.93    & \nodata \\  \hline
 030406       &  4.81 &  4.81 &  4.26 &  1.62    & \nodata           &  0.40    &   1.3   \\  \hline
 030519B      & 10.27 & 10.36 &  9.56 &  6.07    & $6.10 \pm 0.1$    &  1.78    & \nodata \\  \hline 
 031027       &  5.37 &  5.45 &  4.81 &  3.97    & \nodata           &  1.17    &   1.4   \\  \hline 
 031111       &  7.40 &  7.40 &  6.59 &  2.10    & 1.714             &  0.56    &   0.21 
\enddata     
\tablecomments{
\fluenceCB: fluence from  \CB\ fit (Table~\ref{tab:CB_pars});
\fluenceBand: fluence from Band function fit (Table~\ref{tab:Band_pars});
\fluenceRHESSI: fluence in [100,10000]~\keV\ (RHESSI range);
\fluenceHETE: fluence in [30,400]~\keV\ (HETE range);
\fluenceUlysses: fluence in [25,100]~\keV\ (Ulysses range);
HETE: HETE fluences from references cited in \S \ref{sec:results};
Ulysses: Ulysses fluences from references cited in Table~\ref{tab:GRBs}.
}
\tablenotetext{a}{Our fits overestimate the real counts. More realistic is
\fluenceUlysses$=4.8\times10^{-5}\,\mbox{erg cm}^{-2}$.
}
\end{deluxetable}

\begin{deluxetable}{cccccccc}
\tablecolumns{8}
\tablecaption{Peak resolved analysis of GRB 021206
         \label{tab:dt_grb021206}}
\tablewidth{0pt}
\tablehead{
\colhead{$\Delta t$} &
\colhead{duration} &
\colhead{$T_p$} & 
\colhead{$\alpha$} & 
\colhead{$\beta$} & 
\colhead{$b$} & 
\colhead{$F_{CPL}$} & 
\colhead{$F_{mPL}$}
\\
\colhead{} &
\colhead{(s)} &
\colhead{(\keV)} &
\colhead{} &
\colhead{} &
\colhead{} &
\multicolumn{2}{c}{($10^{-5}\,$erg cm$^{-2}$)}
}
\startdata
P1   & 1.366 & $661 \pm 18$  & $0.77 \pm 0.06$ & $2.1$ & $0.059 \pm 0.010$        & 11.7   & \phn3.0 \\
P2   & 1.366 & $732 \pm 12$  & $0.42 \pm 0.05$ & $2.1$ & $0.115 \pm 0.010$        & 14.5   & 12.9    \\
P3   & 1.366 & $684 \pm 14$  & $0.63 \pm 0.05$ & $2.1$ & $0.085 \pm 0.012$        & 12.7   & \phn6.3 \\
P4   & 1.366 & $530 \pm 20$  & $0.80 \pm 0.08$ & $2.1$ & $0.113 \pm 0.018$        & \phn6.9& \phn3.8 \\ 
tail & 4.097 & $160 \pm 60$  &  1.0            & $2.1$ & $2.5\;^{+\infty}_{-1.5}$ & \phn0.2& \phn3.1
\enddata
\tablecomments{
 $\Delta t$: time period, cf.\ Fig.~\ref{fig:ltc_GRBsI};
 $T_p$, $\alpha$, $\beta$ and $b$: \CB\ parameters;
 \fluenceCPL: fluence of the CPL-component in the \CB\ function (\refeq{eq:CB}),
 \fluencemPL: fluence of the modified PL component of \refeq{eq:CB};
 fluences are for the range [100,10000]~\keV.
}
\end{deluxetable}

\clearpage

\end{document}